\documentclass[aerospace,article,submit,pdftex,moreauthors]{Definitions/mdpi}

\firstpage{1} 
\pubvolume{1}
\issuenum{1}
\articlenumber{0}
\pubyear{2023}
\copyrightyear{2023}
\datereceived{} 
\dateaccepted{} 
\datepublished{}
\hreflink{https://doi.org/} 

\usepackage[utf8]{inputenc}
\usepackage[T1]{fontenc}
\usepackage{wrapfig}
\usepackage{multirow}
\usepackage{booktabs}       
\usepackage{amsfonts}       
\usepackage{amsmath}
\usepackage{nicefrac}       
\usepackage{microtype}      

\usepackage{pdfpages}
\usepackage{bbding}

\usepackage[normalem]{ulem} 

\definecolor{Gray}{gray}{0.9}
\definecolor{Pink}{HTML}{FFA07A} 
\definecolor{Blue}{HTML}{728ed6} 

\Title{A Virtual Simulation-Pilot Agent for Training of Air Traffic Controllers}

\TitleCitation{A Virtual Simulation-Pilot Agent for Training of Air Traffic Controllers}

\Author{Juan Zuluaga-Gomez $^{\star, 1,2}$\orcidA{}, Amrutha Prasad$^{1,4}$, Iuliia Nigmatulina$^{1,3}$, Petr Motlicek$^{1,4}$\orcidB{}, Matthias Kleinert$^{5}$ 
}

\AuthorNames{Juan Zuluaga-Gomez, Amrutha Prasad, Iuliia Nigmatulina, Petr Motlicek, Matthias Kleinert}

\address{$^{1}$ \quad Idiap Research Institute, Martigny, Switzerland\\
$^{2}$ \quad Ecole Polytechnique F\'ed\'erale de Lausanne, Switzerland \\
$^{3}$ \quad University of Zurich, Switzerland\\
$^{4}$ \quad Brno University of Technology, Brno, Czechia\\
$^{5}$ \quad German Aerospace Center (DLR), Institute of Flight Guidance, Braunschweig, Germany \\
}

\corres{Correspondence: juan-pablo.zuluaga@idiap.ch}

\abstract{
In this paper we propose a novel virtual simulation-pilot engine for speeding up air traffic controller (ATCo) training by integrating different state-of-the-art artificial intelligence (AI) based tools.
The virtual simulation-pilot engine receives spoken communications from ATCo trainees, and it performs automatic speech recognition and understanding. Thus, it goes beyond only transcribing the communication and can also understand its meaning. The output is subsequently sent to a response generator system, which resembles the spoken read back that pilots give to the ATCo trainees.
The overall pipeline is composed of the following submodules:
(i) automatic speech recognition (ASR) system that transforms audio into a sequence of words; 
(ii) high-level air traffic control (ATC) related entity parser that understands the transcribed voice communication; and (iii) a text-to-speech submodule that generates a spoken utterance that resembles a pilot based on the situation of the dialogue. 
Our system employs state-of-the-art AI-based tools such as Wav2Vec 2.0, Conformer, BERT and Tacotron models. To the best of our knowledge, this is the first work fully based on open-source ATC resources and AI tools. 
In addition, we have developed a robust and modular system with optional submodules that can enhance the system's performance by incorporating real-time surveillance data, metadata related to exercises (such as sectors or runways), or even introducing a deliberate read-back error to train ATCo trainees to identify them. Our ASR system can reach as low as 5.5\% and 15.9\% word error rates (WER) on high and low-quality ATC audio. We also demonstrate that adding surveillance data into the ASR can yield callsign detection accuracy of more than 96\%.   
}


\begin{document}

\section{Introduction}
\label{sec:introduction}

The exponential advances in artificial intelligence (AI) and machine learning (ML) have opened the door of automation to many applications. Examples are automatic speech recognition (ASR)~\cite{nassif2019speech} applied to personal assistants (e.g., SIRI\textregistered~or Amazon's ALEXA\textregistered) and natural language processing (NLP) and understating~\cite{otter2020survey} for different tasks such as sentiment analysis~\cite{zhang2018deep} and user intent detection~\cite{Lugosch2019}. Even though these advances are remarkable, many applications have lagged behind due to their critical matter, imperative near-to-perfect performance or simply because the users or administrators only trust the already existent legacy systems. One clear example is air traffic control (ATC) communications. 

In ATC communications, ATCos are required to issue verbal commands\footnote{Although different means of communication such as CPDLC: Controller Pilot Data Link Communications (CPDLC). CPDLC is a two-way data-link system by which controllers can transmit non-urgent strategic messages to an aircraft as an alternative to voice communications. The message is displayed on a flight deck visual display.} to pilots in order to keep control and safeness of a given area of airspace. The communication protocol has remained almost unchanged since the last century. 

\begin{figure}[t!]
    \centering
    \includegraphics[width=0.75\linewidth]{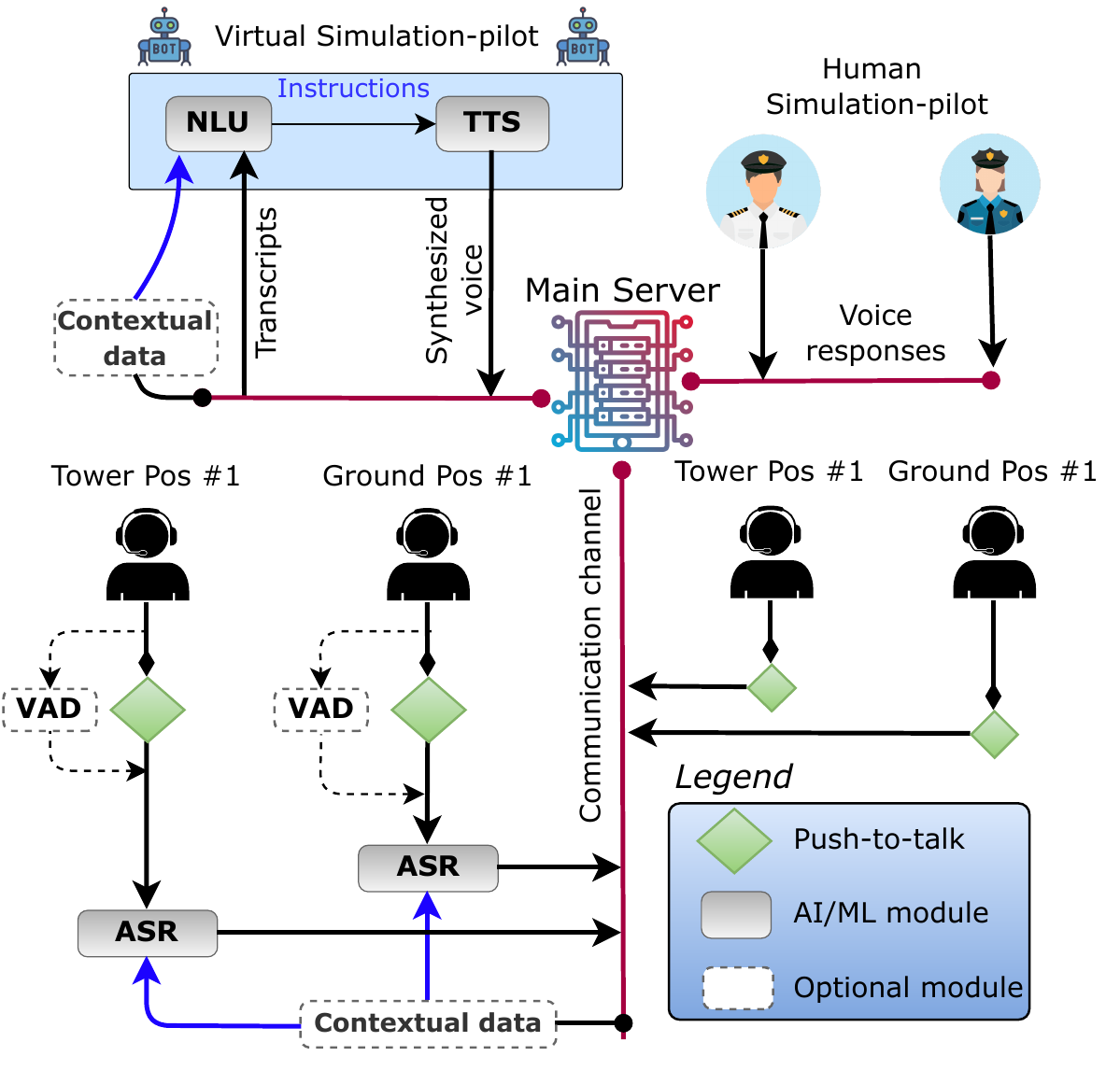}
    \caption{\textit{Virtual simulation-pilot pipeline for ATCo training.} A traditional ATCo training setup is depicted on the right side, while our proposed virtual simulation-pilot is on the left side. The pipeline starts with an ATCo trainee issuing a communication and its capture after the end of the PTT signal (or voice activity detection if the PTT signal is not available). Then, the ASR and \textit{\textbf{high-level entity parser}} (NLP) modules transcribe and extract the ATC-related entities from the voice communication. The output is later rephrased with `simulation-pilot' grammar. The speech synthesizer uses the generated text to create a WAV file containing the spoken textual prompt. In the end, a response is generated by the virtual simulation-pilot, that matches the desired read back.}
    \label{fig:full_pipeline}
\end{figure}

Research targeted at understanding spoken ATC communications in the military domain can be traced back to the 70s~\cite{beek1977}, late 80s~\cite{hamel1989}, and 90s~\cite{matrouf1990adapting,weinstein1991opportunities}. 
Recent projects\footnote{Examples are:  MALORCA~\cite{kleinert2018semi,kleinert2019adaptation,helmke2016reducing,helmke2017increasing}, STARFish~\cite{kleinert2021apron}, HAAWAAI~\cite{nigmatulina2022two,kleinert2021automated} and ATCO2~\cite{zuluaga2022atco2,kocour2021automatic,kocour21_interspeech}. These have shown mature-enough ASR and NLP systems that demonstrate potential for deployment in real-life operation control rooms.}
are aiming at integrating AI-based tools by (i) developing robust acoustic-based AI systems for transcribing dialogues. Examples are voice activity detection (VAD), diarization~\cite{zuluaga2021bertraffic} and ASR~\cite{lin2021improving,zuluaga2020automatic,de2006air,fan2021speech,enac,vocalise}; (i) a few researchers have gone further by developing techniques to understand the ATCo-pilot dialogues~\cite{helmke2016reducing,helmke2017increasing,zuluaga2022atco2,helmke2018ontology}. However, these previous works are mostly disentangled from each other. Some researchers only focus on ASR, while others on high-level entity extraction, such as callsigns and commands. 

Another key application that has seen growth in interest is ATCo training framework. Training ATCos usually involves a human simulation-pilot. The simulation-pilot responds to or issues a request to the ATCo trainee in order to simulate an ATC communication with standard phraseology~\cite{allclear}. It is a human-intensive task, where a specialized workforce is needed\footnote{A human simulation-pilot is required in EUROCONTROL's ESCAPE lite simulator~(\texttt{\url{https://www.eurocontrol.int/simulator/escape}}) during ATCo training.} and the overall cost is usually high. In a standard training scenario, the default simulation-pilots (humans) are required to execute the steps given by ATCos trainees, as in the case of real pilots (directly introduced to the simulator). The pilots, on the other hand, update the training simulator, so that the ATCos can see whether the pilots are following the desired orders. Therefore, this simulation is very close to a real ATCo-pilot communication.
One well-known tool for ATCo training is Eurocontrol's ESCAPE simulator. It is an air traffic management (ATM) real-time simulation platform that supports: (i)~airspace design for en-route and terminal maneuvering areas; (ii)~evaluation of new operational concepts and ATCo tools; (iii)~pre-operational validation trials; and most importantly, (ii)~the \textbf{training of ATCos}~\cite{bouchal2022design}. On our side, we develop a virtual simulation-pilot engine that understands ATCo trainees' commands and possibly can replace current simulators based on human simulation-pilots\footnote{In practice, the proposed virtual simulation-pilot can handle simple ATC communications, e.g., the first phase of the ATCo trainee's training. Thus, humans are still required for more complex scenarios.}. Analogous efforts of developing a virtual simulation-pilot agent have been covered in~\cite{lin2021spoken,lin2021deep}.

In this paper, we enlarge our previous work presented at SESAR Innovation Days 2022~\cite{prasad2022pseudo_pilot}. In~\cite{prasad2022pseudo_pilot}, we introduce a simple yet efficient `proof-of-concept' virtual simulation-pilot. Here, we formalize our system, add more ATM-related modules, and demonstrate that open-source AI-based models area a good fit for the ATC domain. A contrast between our proposed pipeline (left side) and the current human-based simulation-pilot (right side) approaches for ATCo training is illustrated in Figure~\ref{fig:full_pipeline}.

\textbf{Main contributions:} our work proposes a novel virtual simulation-pilot system based on fine-tuning several open-source AI models with ATC data. Our mains contributions are:

\begin{itemize}
    \item The paper presents an end-to-end pipeline that utilizes a virtual simulation-pilot, capable of replacing human simulation-pilots. Implementing this pipeline can speed up the training process of ATCos while decreasing overall training costs.
    \item The virtual simulation-pilot system is modular, allowing for a wide range of domain-specific contextual data to be incorporated, such as real-time air surveillance data, runways numbers, or sectors proper from the given training exercise. This flexibility helps improve the system performance and makes it easier to adapt it to various simulation scenarios, including adaptation to different airports.
    \item Our pipeline is built entirely on open-source and state-of-the-art pre-trained AI models that have been fine-tuned on the ATC domain. Wav2Vec 2.0 and XLSR~\cite{wav2vec,xlsr} models are used for ASR, BERT~\cite{devlin2018bert} is employed for NLU, and FastSpeech2~\cite{ren2020fastspeech} is used for the TTS module. To the best of our knowledge, this is the first work that utilizes open-source ATC resources exclusively~\cite{LDC_ATCC,zuluaga2022atco2,UWB_ATCC,ATCOSIM}.
    \item The virtual simulation-pilot engine is highly versatile and can be customized to suit any potential use case. For example, the system can employ either a male or female voice, or simulate 'very high-frequency noise' to mimic real-life ATCo-pilot dialogues. Additionally, new rules for NLP and ATC understanding can be integrated based on the target application, such as approach or tower control. 
\end{itemize}

We believe this research is a game changer in the ATM community due to two aspects. First, we introduce a novel modular system that can be adjusted to specific scenarios, e.g., tower, ground control, and area control center. Second, we demonstrate that open-source models like XLSR~\cite{xlsr} (for ASR) or BERT~\cite{devlin2018bert} (for NLP and ATC understanding) can be successfully adapted to the ATC scenario. 
We hope the proposed virtual simulation-pilot engine will be the starting point to develop more inclusive and mature systems aimed at ATCo training.

The rest of the paper is organized as follows.
Section~\ref{sec:virtual-pp} describes the virtual simulation-pilot system. We cover the fundamental background for each of the base (Section~\ref{subsec:base-mod}) and optional modules (Section~\ref{subsec:optional-mod}) of the system. Section~\ref{sec:datasets} describes the databases used. Then, Section~\ref{sec:experiment-setup-results} covers the experimental setup followed for adapting the virtual simulation-pilot and the results for each module of the system. Finally, we provide brief future work directions in~\ref{sec:future-work} and conclude the paper in Section~\ref{sec:conclusion}.

\section{Virtual Simulation-Pilot System}
\label{sec:virtual-pp}

The virtual simulation-pilot system manages the most commonly used commands in ATC (see the list in Table~\ref{tab:word-fixer-mod}). It is particularly well-suited for the early stages of ATCo training. Its modular design allows the addition of more advanced rules and grammar to enhance the system's robustness. Our goal is to enhance the foundational knowledge and skills of ATCo trainees. Furthermore, the system can be customized to specific conditions or training scenarios, such as when the spoken language has a heavy accent (e.g., the case of foreign English) or when the ATCo trainee is practicing different positions e.g., ground control or approach/departure. 

In general, ATC communications play a critical role in ensuring the safe and efficient operation of aircraft. These communications are primarily led by ATCos, who are responsible for issuing commands and instructions to pilots in real-time. The training process of ATCos involves 3 stages: (i) initial, (ii) operational, and (iii) continuation training. The volume and complexity of these communications can vary greatly depending on factors such as airspace conditions and seasonal fluctuations, with ATCos often facing increased workloads during peak travel seasons~\cite{prasad2022pseudo_pilot}. As such, ATCo trainees must be prepared to handle high-stress and complex airspace situations through a combination of intensive training and simulation exercises with human simulation-pilots~\cite{pavlinovic2017air}. In addition to mastering the technical aspects of air traffic control, ATCo trainees must also develop strong communication skills, as they are responsible for ensuring clear and precise communication with pilots at all times.

Due to the crucial aspect of ATM, efforts have been made to develop simulation interfaces for their training~\cite{pavlinovic2017air,jurivcic2011role,chhaya2018enhancing}. The previous work includes optimization of the training process~\cite{updegrove2017optimization}, post-evaluation of each training scenario~\cite{eide2014post,nemethova2019education}, virtual simulation-pilot implementation, for example, a deep learning (DL) based implementation~\cite{lin2021deep,zhang2022automatic}. In~\cite{lin2021deep}, the authors use sequence-to-sequence DL models to map from spoken ATC communications to high-level ATC entities. They use the well-known, transformer architecture~\cite{vaswani2017attention}. Transformer is the base of the recent, well-known, encoder-decoder models for ASR (Wav2Vec 2.0~\cite{wav2vec}) or NLP (BERT~\cite{devlin2018bert}). In the next subsections, we will address in more detail each module that is a part of the virtual simulation-pilot system.  

\subsection{Base Modules}
\label{subsec:base-mod}

The proposed virtual simulation-pilot system (see Figure~\ref{fig:full_pipeline}) is built with a set of base modules, and possibly, optional submodules. The most simple version of the engine contains only the base modules. 

\subsubsection{Automatic Speech Recognition}
\label{subsec:asr}

\begin{figure}[t]
    \centering
    \includegraphics[width=0.95\linewidth]{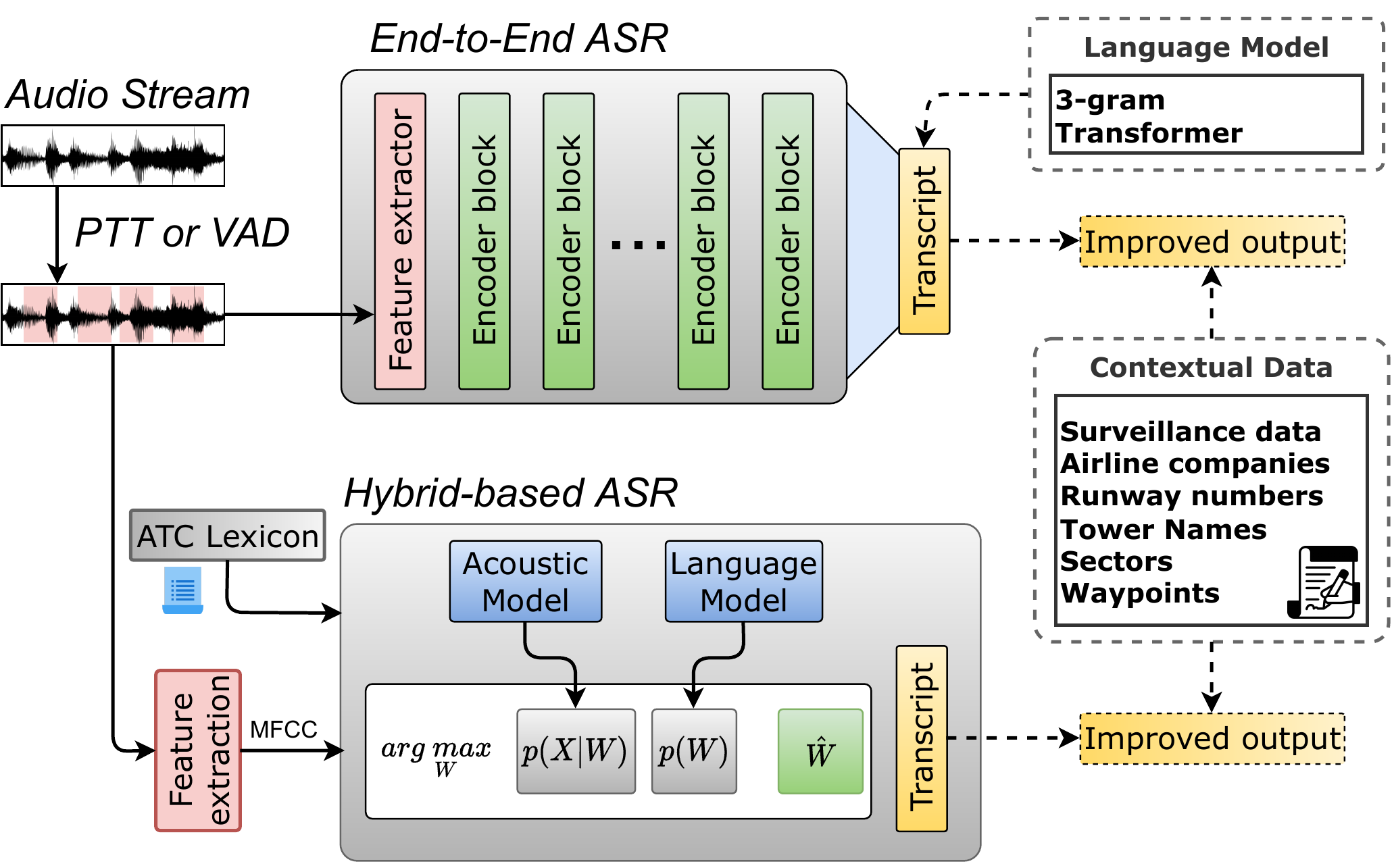}
    \caption{\textit{Traditional hybrid-based and more recent automatic speech recognition systems}. These systems take an ATCo voice communication as input and then produce transcripts as output. The dotted blocks refer to modules that are optional. For instance, surveillance data or other type of data (e.g., sector or waypoints) can be added to increase the overall performance of the system.}
    \label{fig:e2e_asr}
\end{figure}

Automatic Speech Recognition (ASR) or speech-to-text system converts speech to text. An ASR system uses an acoustic model (AM) and a language model (LM). 
The AM represents the relationship between a speech signal and phonemes/linguistic units that make up speech and is trained using the speech recordings along with its corresponding text. 
The LM provides a probability distribution over a sequence of words and provides context to distinguish between words and phrases that sound similar and is trained using a large corpus of text data. A decoding graph is built as a Weighted Finite State Transducer (WFST)~\cite{mohri2002weighted,mohri2008speech,riley2009openfst} using the AM and LM that generates text output given an observation sequence. Standard ASR systems rely on a lexicon, LM, and AM as stated above. Currently, there are two main ASR paradigms, where different strategies, architectures, and procedures are employed for blending all these modules in one "system". The first is hybrid-based ASR, while the second is a more recent approach, termed end-to-end ASR. A comparison of both is drawn in Figure~\ref{fig:e2e_asr}. \\

\textbf{Hybrid Based Automatic  Speech Recognition:} ASR with hybrid systems is based on Hidden Markov Models (HMM) and Deep Neural Networks (DNN)~\cite{dahl2011context,vesely2013sequence}. DNNs are an effective module for estimating the posterior probability of a given set of possible outputs (e.g., phone-state or tri-phone state probability estimator in ASR systems). These posterior probabilities can be seen as pseudo-likelihoods or “scale likelihoods”, which can be interfaced with HMM modules. HMMs provide a structure for mapping a temporal sequence of acoustic features, $\boldsymbol{X}$, e.g., Mel-frequency cepstral coefficients (MFCCs) into a sequence of states \cite{morgan1993hybrid,bourlard1993connectionist}. Hybrid systems remain one of the best approaches for building ASR engines. Currently, HMM-DNNs based ASR is the state-of-the-art systems for ASR in ATC domain~\cite{srinivasamurthy2017semi,kleinert2018semi,zuluaga2020automatic}. 

Recent work in ASR has targeted different areas in ATC. For instance, a benchmark for ASR on ATC communications databases is established in~\cite{zuluagagomez20_interspeech}. Leveraging non-transcribed ATC audio data by semi-supervised learning has been covered in~\cite{srinivasamurthy2017semi,kleinert2018semi,zuluagagomez21_interspeech}. Previous work related to the large-scale automatic collection of ATC audio data from different airports worldwide is covered in~\cite{kocour2021automatic,zuluaga2020automatic}. Additionally, innovative research aimed at improving callsign recognition by integrating surveillance data into the pipeline is covered by~\cite{kocour21_interspeech,nigmatulina2021improving,nigmatulina2022two}.

The main components of a hybrid system are: a pronunciation lexicon, LM, and AM. 
One key advantage of a hybrid system versus other ASR techniques is that the text data (e.g., words, dictionary) and pronunciation of new words are collected and added beforehand, hoping to match the target domain of the recognizer. Standard hybrid-based ASR approaches still rely on word-based lexicons, i.e., missing or out-of-vocabulary words from the lexicon cannot be hypothesized by the ASR decoder. The system is composed of an explicit acoustic and language model. A visual example of hybrid-based ASR systems is in the bottom panel of Figure~\ref{fig:e2e_asr}. \\

\textbf{End-to-End Automatic Speech Recognition:} end-to-end (E2E) systems are based on a  different paradigm compared to hybrid-based ASR. E2E-ASR aims at directly transcribing speech to text without requiring alignments between acoustic frames (i.e., input features) and output characters/words, which is a necessary separate component in standard hybrid-based systems. Unlike the hybrid approach, E2E models are learning a direct mapping between acoustic frames and model label units (characters, subwords, or words) in one step toward the final objective of interest. 
    
Recent work on encoder-decoder ASR can be categorized into two main approaches: Connectionist Temporal Classification (CTC)~\cite{graves2014towards} and attention-based encoder-decoder systems ~\cite{chorowski2015attention}. First, CTC uses intermediate label representation, allowing repetitions of labels and occurrences of `blank output', which labels an output with `no label'. Second, attention-based encoder-decoder or only-encoder models directly learn a mapping from input acoustic frames to character sequences. For each time step, the model emits a character unit conditioned on the inputs and the history of the produced outputs. Important line of works for E2E-ASR can be categorized as self-supervised learning~\cite{schneider2019wav2vec} for speech representation, covering bidirectional models~\cite{wav2vec,chen2021wavlm} or autoregressive models~\cite{oord2018representation, baevski2019vq}. 

Moreover, recent innovative research on E2E-ASR for the ATC domain is covered in~\cite{zuluaga2022does}. Here, the authors follow the practice of fine-tuning a Wav2Vec~2.0 model~\cite{wav2vec} with public and private ATC databases. This system reaches on-par performances with hybrid-based ASR models, demonstrating that this new paradigm for ASR development, also performs well in the ATC domain. In E2E-ASR, the system encodes directly an acoustic and language model, and it produces transcripts in an E2E manner. A visual example of an only-encoder E2E-ASR system is in the top panel of Figure~\ref{fig:e2e_asr}.

\begin{figure}[t]
    \centering
    \includegraphics[width=0.9\linewidth]{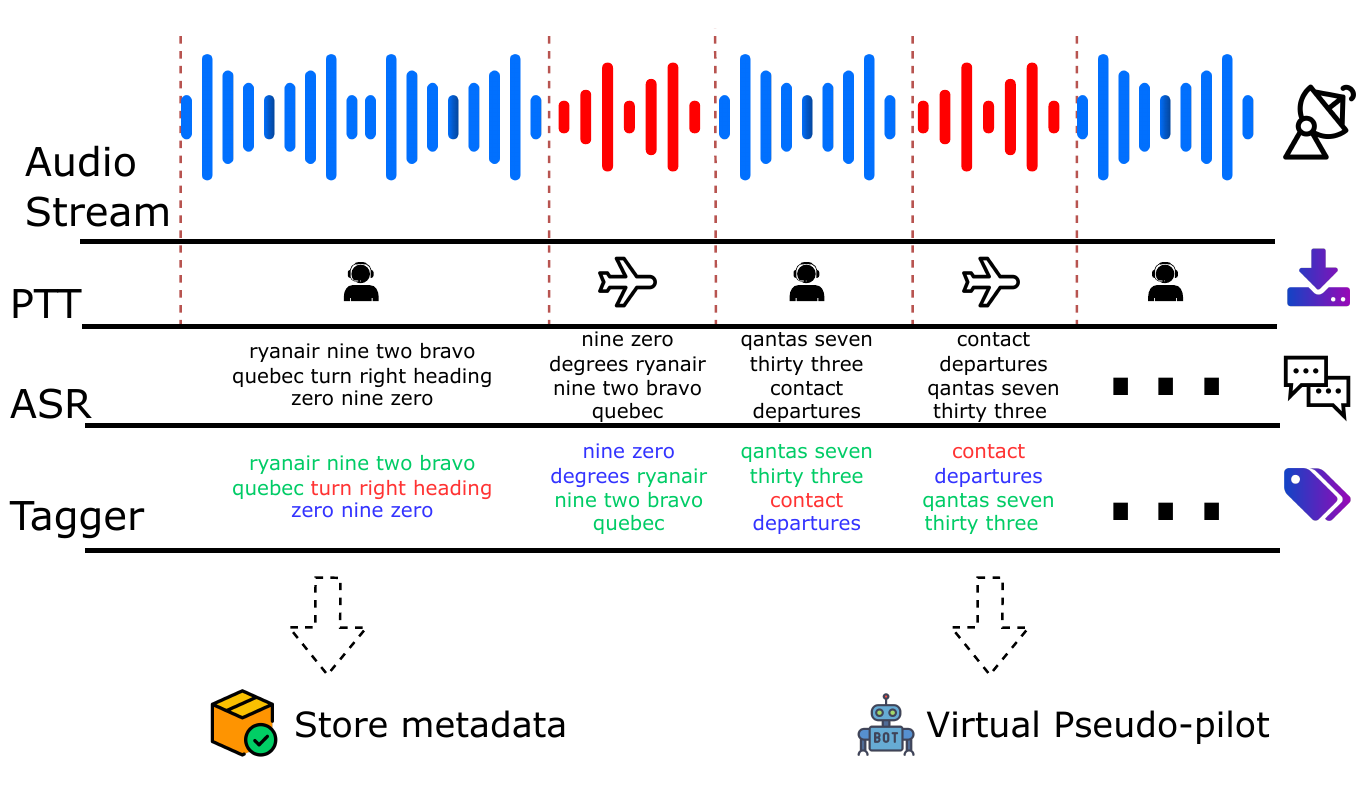}
    \caption{\textit{Detailed outputs of the main ML-based submodules of our proposed simulation-pilot system}. It includes pre-processing from the input audio stream, speaker role detection by push-to-talk (PTT) signal, transcripts generation and callsign/command/value extraction with the high-level entity with ASR and NER modules, respectively. All the data is later aggregated, packaged, and sent to the response generator and TTS module. Note that this data can also be logged into a database for control and recording. Figure adapted from our previous work~\cite{prasad2022pseudo_pilot}.}
    \label{fig:metadata_pipeline}
\end{figure}

\subsubsection{Natural Language Understanding}
\label{subsec:nlu}

Natural language understanding (NLU) is a field of NLP, which aims at reading comprehension. In the field of ATC, NLU is related to intent detection and slot filling\footnote{The slot filling task is akin to named entity recognition (NER).}. In intent detection, we aim at extracting the commands from the communication, while slot filling refers to the 'values' of these commands and callsigns. Further in the paper, we will call the system that extracts the high-level ATC-related knowledge from the ASR outputs as \textit{\textbf{high-level entity parser}} system. The NER-based understanding of ATC communications has been previously studied in~\cite{lin2021spoken,lin2021deep,zuluaga2022atco2}, while our earlier work~\cite{prasad2022pseudo_pilot} includes the integration of named-entity recognition (NER) into the virtual simulation-pilot framework. 

The \textit{\textbf{high-level entity parser}} system is responsible for identifying, categorizing, and extracting crucial keywords and phrases from ATC communications. 
In NLP, these keywords are classified into pre-defined categories such as parts of speech tags, locations, organizations, or individuals' names. In the context of ATC, the key entities include callsigns, commands, values (which includes units, e.g., flight level). For instance, consider the following transcribed communication (taken from Figure~\ref{fig:metadata_pipeline}):

\vspace{0.2cm}

\noindent \textbf{ASR transcript:} \textcolor{black}{\dashuline{ryanair nine two bravo quebec turn right heading zero nine zero}}, 

\vspace{0.2cm} 

\noindent \textbf{would be parsed to high-level ATC entity format:}

\vspace{0.2cm} 

\noindent \textbf{Output:} \textcolor{teal}{<callsign> \dashuline{ryanair nine two bravo quebec} </callsign>} 
\textcolor{red}{<command> \dotuline{turn right heading} </command>} 
\textcolor{blue}{<value> \uwave{zero nine zero} </value>}. 
 
\vspace{0.2cm}

The previous output is then used for further processing tasks. E.g., generate a simulation-pilot-alike response, metadata logging, and reporting, or simply help ATCos in their daily tasks. 

Hereby, regarding NLU, we mostly focus on NER~\cite{yadav2018survey,sharma2022named}. Initially, NER relied on the manual crafting of dictionaries and ontologies, which led to complexity and human error when scaling to more entities or adapting to a different domain~\cite{grishman1996message}. The advancement of ML-based methods for text processing, including NER, has been introduced by~\cite{collobert2011natural}. Previous work such as the one from~\cite{piskorski2017first,yadav2018survey}, continued to advance NER techniques. A \textit{\textbf{high-level entity parser}} system (like ours) can be implemented by fine-tuning a pre-trained LM for the NER task. Currently, state-of-the-art NER models utilize pre-trained LMs such as BERT~\cite{devlin2018bert}, RoBERTa~\cite{liu2019roberta}, or DeBERTa~\cite{he2021deberta}. For the proposed virtual simulation-pilot, we use a fine-tuned BERT on ATC text data.

\subsubsection{Response Generator}
\label{subsec:response-generator}

\begin{figure}[t]
    \centering
    \includegraphics[width=0.9\linewidth]{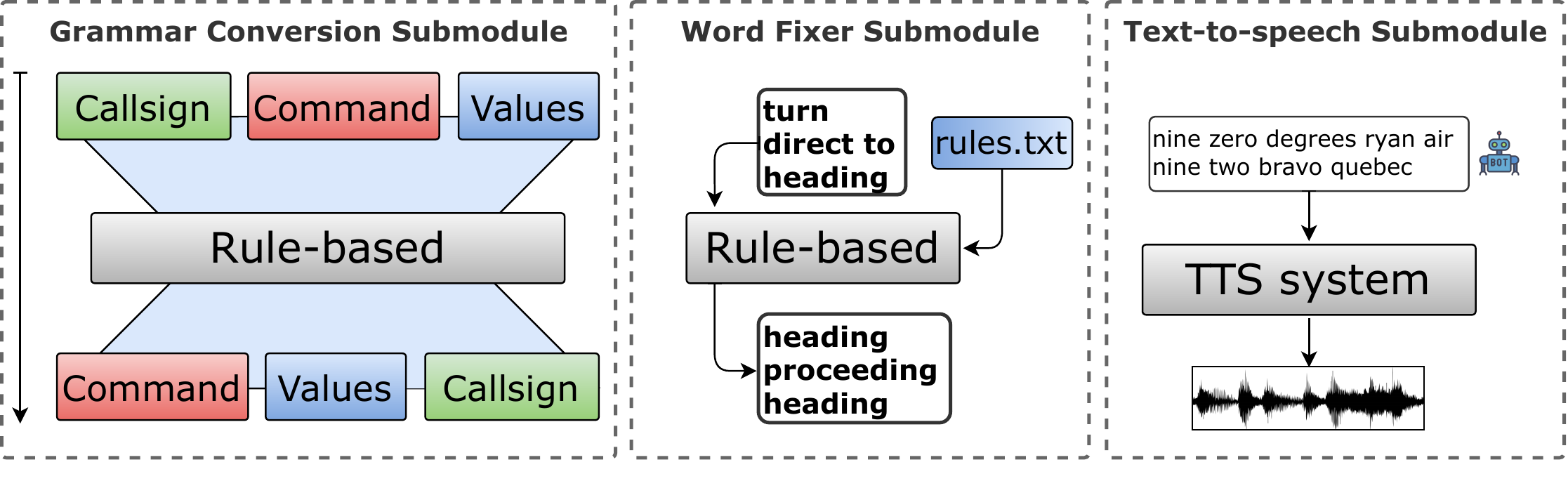}
    \caption{\textit{Detailed submodules of the response generator}.}
    \label{fig:response-generator}
\end{figure}

The response generator (RG) is a crucial component of the simulation-pilot agent. It processes the output from the \textit{\textbf{high-level entity parser}} system, which includes the callsign, commands, and values uttered by the ATCo, and then later generates a spoken response. The response is then delivered in the form of a WAV file, which is played through the headphones of the ATCo trainee. Additionally, the response, along with its metadata, can be stored for future reference and evaluation. The RG system is designed to generate responses that are grammatically consistent with what a standard simulation-pilot (or pilot) would say in response to the initial commands issued by the ATCo. The RG system comprises three submodules: (i) grammar conversion, (ii) word fixer (e.g., ATCo-to-pilot phrase fixer), and (iii) text-to-speech, also known as a speech synthesizer. A visual representation of the RG system split by submodules is in Figure~\ref{fig:response-generator}. \\

\textbf{Grammar Conversion Submodule:} a component designed to generate the response of the virtual simulation-pilot. First, the output of the \textit{\textbf{high-level entity parser}} module (discussed in Section~\ref{subsec:nlu}) is input to the Grammar Conversion Submodule. At this stage, the communication knowledge has already been extracted, including the callsign, commands, and their values. We then perform a grammar adjustment process, rearranging the order of these high-level entities. For example, we take into account the common practice of pilots mentioning the callsign at the end of the utterance while ATCos mention it at the beginning of the ATC communication. Thus, our goal is to align the grammar used by the simulation-pilot with the communication style used by the ATCo. See first left panel in Figure~\ref{fig:response-generator}. \\

\textbf{Word Fixer Submodule:} this is a crucial component of the virtual simulation-pilot system that ensures that the output from the Response Generator aligns with the standard ICAO phraseology. This is achieved by modifying the commands to match the desired response based on the input communication from the ATCo. The submodule applies specific mapping rules, such as converting \textcolor{blue}{\mbox{\textit{descend} $\rightarrow$ \textit{descending}}} or \textcolor{blue}{\mbox{\textit{turn} $\rightarrow$ \textit{heading}}}, to make the generated reply as close to standard phraseology as possible. Similar efforts have been covered in a recent work~\cite{zhang2022automatic} where the authors propose a \textit{'copy mechanism'}, which copies the key entities from the ATCo communication into the desired response of the virtual simulation-pilot, e.g., \textcolor{blue}{\textit{maintain} $\rightarrow$ \textit{maintaining}}. In real-life ATC communication, however, the wording of ATCos and pilots slightly differs. Currently, our \textit{Word Fixer} submodule contains a list of 19 commands but can be easily updated by adding additional mapping rules to a \texttt{rules.txt} file. This allows the system to adapt to different environments, such as ground tower, control departure/approach, or area control center. The main conversion rules used by the Word Fixer submodule are listed in Table~\ref{tab:word-fixer-mod}. The ability to modify and adapt the Word Fixer submodule makes it a versatile tool for training ATCos to recognize and respond to standard ICAO phraseology. See center panel in Figure~\ref{fig:response-generator}. \\

\textbf{Text to Speech Submodule:} speech synthesis, also referred to as text-to-speech (TTS), is a multidisciplinary field that combines various areas of research such as linguistics, speech signal processing, and acoustics. The primary objective of TTS is to convert text into an intelligible speech signal. Over the years, numerous approaches have been developed to achieve this goal, including formant-based parametric synthesis~\cite{klatt1987review}, waveform concatenation~\cite{murray1996emotional}, and statistical parametric speech synthesis~\cite{tokuda2013speech}. In recent times, the advent of deep learning has revolutionized the field of TTS. Models like Tacotron~\cite{Wang2017tacotron} and Tacotron2~\cite{shen2018tacotron2} are end-to-end generative TTS systems that can synthesize speech directly from text input (e.g., characters or words). Most recently, FastSpeech2~\cite{ren2020fastspeech} has gained widespread recognition in the TTS community due to its simplicity and efficient non-autoregressive manner of operation. Finally, TTS is a complex field that draws on a variety of areas of research and has made significant strides recently, especially with the advent of deep learning. For a more in-depth understanding of the technical aspects of TTS engines, we recommend~\cite{He2019,kaur2022conventional} and novel Difussion-based TTS systems in~\cite{jeong2021diff}. See first right panel in Figure~\ref{fig:response-generator}.

\begin{table}[t]
    \centering
    \caption{Word fixing rules. The rules are used to convert ATCos input communications into a virtual simulation-pilot response.}
    \label{tab:word-fixer-mod}
    \begin{tabular}{ l|l }
        \toprule
        \multicolumn{2}{c}{\cellcolor{Gray} \textbf{Word Fixer Submodule - rules.txt}} \\
        \midrule
        \rowcolor{Gray} \multicolumn{1}{c|}{\textbf{\centering Horizontal commands}} & \multicolumn{1}{c}{\textbf{Handover commands}} \\
        \midrule
        continue heading $\rightarrow$ continuing altitude & contact tower $\rightarrow$ contact tower \\
        heading $\rightarrow$ heading & station radar $\rightarrow$ station radar \\
        turn $\rightarrow$ heading & squawk $\rightarrow$ squawk \\
        turn by $\rightarrow$ heading & squawking $\rightarrow$ squawk \\
        direct to $\rightarrow$ proceeding direct & contact frequency $\rightarrow$ NONE \\
        \midrule
        \rowcolor{Gray} \multicolumn{1}{c|}{\textbf{Level commands}} & \multicolumn{1}{c}{\textbf{Speed commands}} \\
        \midrule
        maintain altitude $\rightarrow$ maintaining altitude & reduce $\rightarrow$ reducing \\
        maintain altitude $\rightarrow$ maintain & maintain speed $\rightarrow$ maintaining \\
        descend $\rightarrow$ descending & reduce speed $\rightarrow$ reduce speed \\
        climb $\rightarrow$ climbing & speed $\rightarrow$ NONE \\
        altitude $\rightarrow$ steady & - \\
        \bottomrule
    \end{tabular}
\end{table}

\subsection{Optional Modules}
\label{subsec:optional-mod}

Differently from the base modules, covered in Section~\ref{subsec:base-mod}, the optional modules are blocks that can be integrated into the virtual simulation-pilot to enhance or add new capabilities. An example is the PTT (`push-to-talk') signal. In some cases PTT signal is not available, thus, voice activity detection can be integrated. Below, we cover in more detail each of the proposed optional modules.

\subsubsection{Voice Activity Detection}

Voice Activity Detection (VAD) is an essential component in standard speech processing systems to determine which portions of an audio signal correspond to speech and which are non-speech, i.e. background noise or silence. VAD can be used for offline purposes decoding, as well as for online streaming recognition. The offline VAD is used to split a lengthy audio to shorter segments that can then be used for training or evaluating ASR or NLU systems. The online VAD is particularly crucial for ATC ASR when the PTT signal is not available. An example of an online VAD is the WebRTC\footnote{Developed by Google \url{https://webrtc.org/}}.
In ATC communications, VAD is used to filter out the background noise and keep only speech segments that carry the ATCos (or pilots) voice messages. One of the challenges for VAD in ATC communications is the presence of a high level of background noise. The noise comes from various sources, e.g., engines of aircraft, wind, or even other ATCos. ATC communications can easily have signal-to-noise (SNR) ratios lower than 15 dB. If VAD is not applied (and there is not a PTT signal available), the ASR system may degrade the accuracy of speech transcription, which may result in incorrect responses from the virtual simulation-pilot agent.

VAD has been explored before in the framework of ATC~\cite{sarfjoo2020speech}. A general overview of recent VAD architecture and research directions is covered~\cite{ariav2019end}. Some other researchers have targeted how to personalize VAD system~\cite{ding2019personal}, or how this module plays its role in the framework of diarization~\cite{medennikov2020target}. There are several techniques used for VAD, ranging from traditional feature-based~\cite{Zazo_2016} models, Hidden Markov model, or Gaussian Mixture model based~\cite{ng2012developing}. On the other hand, machine learning-based have proven to be more accurate and robust, particularly deep neural network-based methods. These techniques can learn complex relationships between the audio signal and speech and can be trained on large annotated datasets. For instance, convolutional and deep neural network-based VAD has gotten a lot of interest~\cite{chang2018temporal,sarfjoo2020speech}. VAD can be used in various stages of the ATC communication pipeline. VAD can be applied at the front-end of the ASR system to pre-process the audio signal and reduce the processing time of the ASR system. Figure~\ref{fig:full_pipeline} and Figure~\ref{fig:e2e_asr} depict where a VAD module can be integrated into the virtual simulation-pilot agent.

\subsubsection{Contextual Biasing with Surveillance Data}
\label{context_biasing}

In order to enhance the accuracy of an ASR system's predictions, it is possible to use additional context information along with speech input. In the ATC field, radar data can serve as context information, providing a list of unique identifiers for aircraft in the airspace called "callsigns". By utilizing this radar data, the ASR system can prioritize the recognition of these registered callsigns increasing the likelihood of correct identification. Callsigns are typically a combination of letters, digits, and an airline name, which are translated into speech as a sequence of words. The lattice, or prediction graph, can be adjusted during decoding by weighting the target word sequences using the Finite State Transducer (FST) operation of composition \cite{nigmatulina2021improving,kocour21_interspeech}. This process, called lattice rescoring, has been found to improve recognition accuracy, particularly for callsigns. Multiple experiments using ATC data have demonstrated the effectiveness of this method, especially in improving the accuracy of callsign recognition. The results of contextual biasing are presented and discussed below in section~\ref{sec:asr} and Table~\ref{tab:nats-boosting}. \\

\textbf{Re-ranking module based on Levenshtein distance}: the \textbf{\textit{high-level entity parser}} system for NER (see Section~\ref{subsec:nlu}) allows us to extract the callsign from a given transcript or ASR 1-best hypotheses. Recognition of this entity is crucial where a single error produced by the ASR system affects the whole entity (normally composed of three to eight words). Additionally, speakers regularly shorten callsigns in the conversation making it impossible for an ASR system to generate the full entity (e.g., \textcolor{red}{\textit{`three nine two papa'}} instead of \textcolor{blue}{\textit{`austrian three nine two papa'}}, \textcolor{red}{\textit{`six lima yankee'}} instead of \textcolor{blue}{\textit{`hansa six lima yankee'}}). One way to overcome this issue is to re-rank entities extracted by the \textbf{\textit{high-level entity parser}} system with the surveillance data. The output of this system is a list of tags that match words or sequences of words in an input utterance. As our only available source of contextual knowledge are callsigns registered at a certain time and location, we extract callsigns with the \textbf{\textit{high-level entity parser}} system and discard other entities. Correspondingly, each utterance has a list of callsigns expanded into word sequences.
As input, the re-ranking module takes (i)~a callsign extracted by the \textbf{\textit{high-level entity parser}} system and (ii)~an expanded list of callsigns. The re-ranking module compares a given n-gram sequence against a list of possible n-grams and finds the closest match from the list of surveillance data based on the \textit{weighted Levenshtein distance}. In order to use contextual knowledge, it is necessary to know which words in an utterance correspond to a desired entity (i.e., a callsign), which is why it is necessary to add into the pipeline the \textbf{\textit{high-level entity parser}} system.
We skip the re-ranking in case the output is a `NO\_CALLSIGN' flag (no callsign recognized).

\begin{figure}[t]
    \centering
    \includegraphics[width=0.5\linewidth]{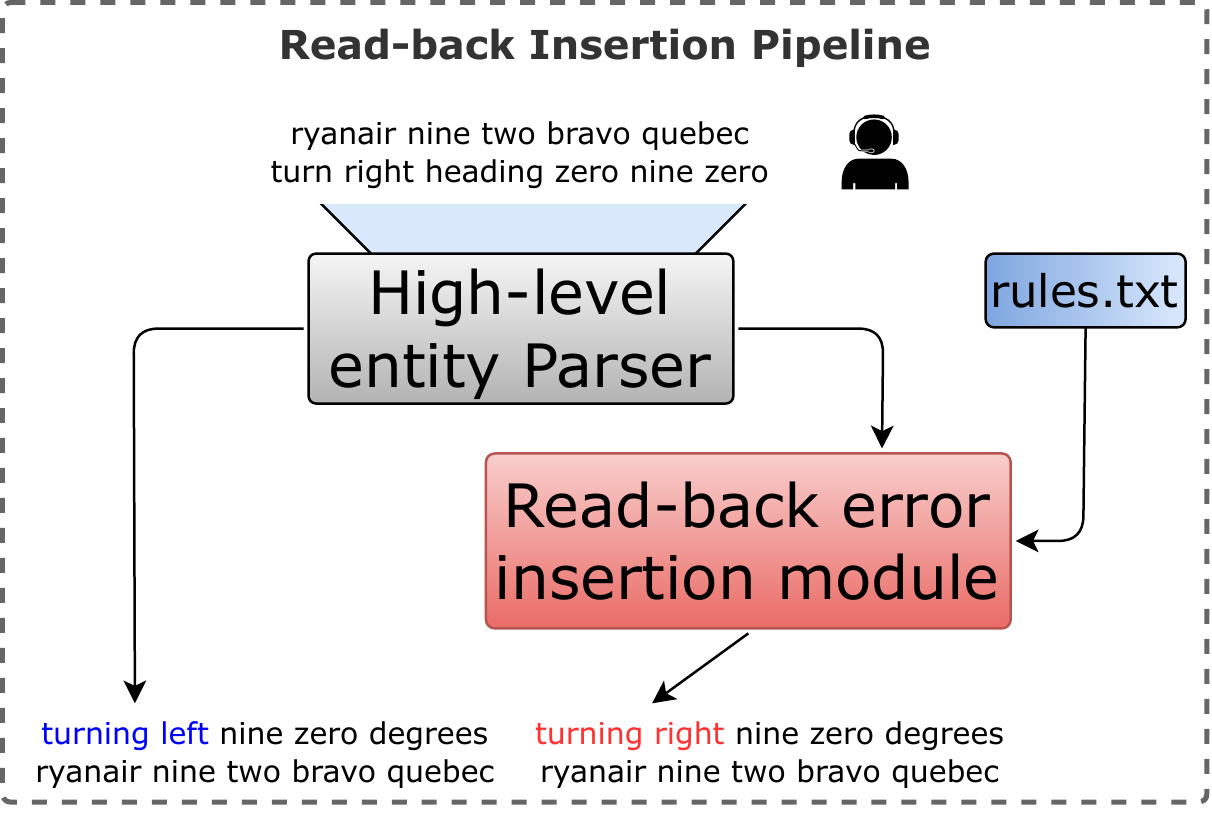}
    \caption{\textit{Read-back insertion module}. At first, an input transcribed communication is sent to the \textit{\textbf{high-level entity parser}} in order to extract the knowledge, i.e., callsign, commands and values. Later, with a defined probability, a desired read-back error can be inserted}
    \label{fig:rbe-insertion-module}
\end{figure}

\subsubsection{Read-back Error Insertion Module}

The approach of using the virtual simulation-pilot system can be adapted to meet various communication requirements in ATC training. This includes creating a desirable read-back error (RBE), which is a plausible scenario in ATC, where a pilot or ATCo misreads or misunderstands a message~\cite{hartmut2022readback}. By incorporating this scenario in ATCos' training, they can develop the critical skills for spotting these errors. This is a fundamental aspect of ensuring the safety and efficiency of ATM~\cite{cordero2012automated}. The ability to simulate (by inserting a desired error) and practice these scenarios through the use of the virtual simulation-pilot system offers a valuable tool for ATCo training and can help to improve the overall performance of ATC. An example could look like: \textcolor{blue}{\mbox{\texttt{ATCo: turn right} $\rightarrow$ \texttt{Pilot (RBE): turning left}}}. 

The structure of the generated RBE could depend on the status of the exercise. For instance, whether the ATCo trainee is in the Ground Tower or Approach/Departure position. These positions should, in the end, change the behavior of this optional module. The proposed, optional RBE insertion module is depicted in Figure~\ref{fig:rbe-insertion-module}.

\section{Datasets}
\label{sec:datasets}

This section describes the public databases used for training and evaluating different modules of our virtual simulation-pilot system. In addition, Table~\ref{tab:databases} summarizes well-known private and public ATC-related databases~\cite{kleinert2021automated,kleinert2018semi,AIRBUS}. Our goal is to conduct a thorough and comprehensive study on the use of virtual simulation-pilots in ATC. To ensure the reproducibility of our research, we use either only open-source databases or a combination of these and private databases during the training phase of our base models\footnote{The exception is the TTS module. Here, we only use a pre-trained out-of-the-box TTS module downloaded from HuggingFace. TTS is part of the response generator.}. Despite this, we are still aiming at demonstrating the potential of our approach in a more realistic setting. Hence, we also report results on private databases that the authors have access to. These results can provide a better idea of the performance of our approach in practical applications, while also highlighting its strengths and weaknesses in a real-world scenario. In any case, our focus remains on ensuring that our research is thoroughly documented and that it can be easily replicated by other researchers in the ATC and ATM field.

\subsection{Public databases}
\label{subsec:pub-databases}

\textbf{LDC-ATCC corpus:} the Air Traffic Control Corpus\footnote{LDC-ATCC: \texttt{\url{https://catalog.ldc.upenn.edu/LDC94S14A}}.} (ATCC) consists of recorded speech initially designed for research on ASR. However, we employ the recordings along the metadata for NLU research, e.g., speaker role detection. The audio data contains voice communication traffic between various ATCos and pilots. The audio files are sampled at 8 kHz, 16-bit linear, representing continuous monitoring without squelch or silence elimination. Each file has a single frequency over one to two hours of audio. The corpus contains gold annotations and metadata.\footnote{Metadata covers voice activity segmentation details, speaker role information (who is talking), and callsigns in ICAO format.}
The corpus consists of approximately 25\,h of ATCo and pilot transmissions (after VAD).

\textbf{UWB-ATCC corpus:} the UWB-ATCC\footnote{Corpus released by the University of West Bohemia: \texttt{\url{https://lindat.mff.cuni.cz/repository/xmlui/handle/11858/00-097C-0000-0001-CCA1-0}}.} corpus is a free and public resource for research on ATC. It contains recordings of communication between ATCos and pilots. The speech is manually transcribed and labeled with the speaker information, i.e., pilot/controller. The total amount of speech after removing silences is 13\,h. The audio data is mono-channel sampled at 8kHz and 16-bit PCM.

\textbf{ATCO2 corpus:} dataset built for the development and evaluation of ASR and NLP technologies for English ATC communications. The dataset consists of English coming from several airports worldwide (e.g., LKTB, LKPR, LZIB, LSGS, LSZH, LSZB, YSSY). 
We used this corpus twofold. First, we employ up to 2500 hours of audio data of the official pseudo-annotated set (see more information in~\cite{zuluaga2022atco2}) for training our ASR systems, this training set is labeled as \textit{ATCO2-PL set corpus}. It is worth mentioning, that the transcripts of the ATCO2 training corpus were automatically generated by an ASR system. Despite that, recent work has shown its potential to develop robust ASR systems for ATC from scratch, e.g.,~\cite{zuluaga2022atco2}. Secondly, for completeness, we use the two official partitions of the ATCO2 test set, namely, \textit{ATCO2 test set 1h corpus} and the \textit{ATCO2 test set 4h corpus} as evaluation sets. The first corpus contains 1.1\,hr of open-sourced transcribed annotations, and it can be accessed for free in \texttt{\url{https://www.atco2.org/data}}. The latter contains $\sim$3\,h of extra annotated data, and the full corpus is available for purchase through ELDA in
\texttt{\url{http://catalog.elra.info/en-us/repository/browse/ELRA-S0484}}. The recordings are mono-channel sampled at 16kHz and 16-bit PCM. 

\textbf{ATCOSIM corpus}: a free public database for research on ATC communications. It comprises 10\,h of speech data recorded during ATC real-time simulations using a close-talk headset microphone. The utterances are in English language and pronounced by ten non-native speakers. The speakers are split by gender. Even though we do not pursue this direction, the ATCOSIM corpus can be used for the development or adaptation of already existent TTS systems to ATC with voices from different gender, i.e., males or females. ATCOSIM also includes orthographic transcriptions and additional information about speakers and recording sessions~\cite{ATCOSIM}. This dataset can be accessed for free: \url{https://www.spsc.tugraz.at/databases-and-tools}.

\begin{table}
    \centering
    \caption{Air traffic control communications related databases. $^{\star}$abbreviation in IETF format. $^\dagger$research directions that can be explored based on the annotations provided by the dataset. 
    $^{\dagger\dagger}$ASR and TTS are interchangeable, the same annotations of each recording can be used to fine-tune or train a TTS module. $^{\mathsection}$denotes datasets that contain annotations on the callsign or/and command level.}
    \label{tab:databases}
    \begin{tabular}{lll ccc cc}
        \toprule
        \multicolumn{3}{c}{\cellcolor{Gray} \textbf{Characteristics}} & \multicolumn{3}{c}{\cellcolor{Gray} \textbf{Research Topics}$^{\dagger}$} &  \multicolumn{2}{c}{\cellcolor{Gray} \textbf{Other}} \\
        \cmidrule(lr){1-3}
        \cmidrule(lr){4-6}
        \cmidrule(lr){7-8}
        Database & Accents$^{\star}$ & Hrs & ASR/TTS$^{\dagger\dagger}$ & SpkID & NLU$^{\mathsection}$ & License & Ref. \\
        \midrule
        \rowcolor{Gray}
        \multicolumn{8}{c}{\textbf{\textit{Private databases}}} \\
        \midrule                
        MALORCA & cs, de & 14 & \textcolor{blue}{\checkmark}  & \textcolor{blue}{\checkmark} & \textcolor{blue}{\checkmark} & \textcolor{brown}{\XSolidBrush} &  \cite{kleinert2018semi,srinivasamurthy2017semi} \\
        AIRBUS & fr & 100 & \textcolor{blue}{\checkmark} & - & \textcolor{blue}{\checkmark} & \textcolor{brown}{\XSolidBrush} &  \cite{AIRBUS}\\
        HAAWAII & is, en-GB & 47 & \textcolor{blue}{\checkmark} &\textcolor{blue}{\checkmark} &\textcolor{blue}{\checkmark} & \textcolor{brown}{\XSolidBrush} & \cite{zuluaga2022does,kleinert2021automated} \\  
        Internal & several & 44 & \textcolor{blue}{\checkmark} &\textcolor{brown}{\XSolidBrush} &\textcolor{brown}{\XSolidBrush} & \textcolor{brown}{\XSolidBrush} & - \\  
        \midrule
        \rowcolor{Gray}
        \multicolumn{8}{c}{\textbf{\textit{Public databases}}} \\
        \midrule                
        ATCOSIM & de, fr, de-CH & 10.7 & \textcolor{blue}{\checkmark} & \textcolor{brown}{\XSolidBrush} & \textcolor{brown}{\XSolidBrush} & \textcolor{blue}{\checkmark} &  \cite{ATCOSIM} \\
        UWB-ATCC & cs & 13.2 & \textcolor{blue}{\checkmark} & \textcolor{blue}{\checkmark} &  \textcolor{brown}{\XSolidBrush} & \textcolor{blue}{\checkmark} & \cite{UWB_ATCC} \\
        LDC-ATCC & en-US & 26.2 &  \textcolor{blue}{\checkmark} & \textcolor{blue}{\checkmark} & \textcolor{blue}{\checkmark} & \textcolor{blue}{\checkmark} & \cite{LDC_ATCC} \\
        HIWIRE & fr, it, es, el & 28.7 & \textcolor{blue}{\checkmark} & \textcolor{brown}{\XSolidBrush} & \textcolor{brown}{\XSolidBrush} & \textcolor{blue}{\checkmark} & \cite{HIWIRE} \\
        ATCO2 & several & 5285 & \textcolor{blue}{\checkmark} & \textcolor{blue}{\checkmark} & \textcolor{blue}{\checkmark} & \textcolor{blue}{\checkmark} &  \cite{zuluaga2022atco2}\\
        \bottomrule
    \end{tabular}
\end{table}

\subsection{Private databases}
\label{subsec:private-databases}

\textbf{MALORCA corpus}\footnote{MAchine Learning Of speech Recognition models for Controller Assistance: \url{http://www.malorca-project.de/wp/}}: is based on the research project that focuses to propose general, cheap and effective solution to develop and automate the speech recognition for controllers using the speech data and contextual information. The data collected is mainly from Prague and Vienna airports which is around 14\,h. The data is split to train and test with a split amount of 10\,h and 4\,h (2\,h from each airport) respectively.

\textbf{HAAWAII corpus}\footnote{Highly Advanced Air Traffic Controller Workstation with Artificial Intelligence Integration: \texttt{\url{https://www.haawaii.de}}.}: dataset based on an exploratory research project that aims to research and develop a reliable and adaptable solution to automatically transcribe voice commands issued by both ATCos and pilots. The controller and pilot conversations are obtained from two Air Navigation Service Providers (ANSPs): (i) NATS for London Approach and (ii) ISAVIA for Icelandic en-route. The total amount of manually transcribed data available is around 47\,h (partitioned into 43\,h for train and 4\,h for test). The 4\,h test set is taken -- 2\,h each -- from both London and Iceland airports. Similar to another corpus, the audio files are sampled at 8\,kHz and 16-bit PCM. This corpus is only used as an out-of-domain dataset, thus we only report results on the ASR level.

\textbf{Internal data}:
In addition to the above mentioned datasets, we have data from some industrial research projects which amounts to a total duration of 44\,h of speech recordings of ATCO and pilot along with their manual transcripts.

\section{Experimental setup \& Results}
\label{sec:experiment-setup-results}

In this section, we present experimental results for some modules described in Section~\ref{sec:virtual-pp}. These modules are trained with the datasets from Section~\ref{sec:datasets}. Note that all datasets are not used during training and testing phases.

\subsection{Automatic Speech Recognition}
\label{sec:asr}

This subsection list the results related to ASR, previously covered in Section~\ref{subsec:asr}. We analyze (i) the three proposed ASR architectures, (ii) the train datasets used during the training phase, and (iii) the experimental setup and results obtained on different public and private test sets. 

\subsubsection{Architectures}

The results of ASR are split in two. First, we evaluate hybrid-based ASR models, which are the default in current ATC-ASR research~\cite{helmke2016reducing,helmke2017increasing}. Second, we train ASR models with state-of-the-art end-to-end architectures, e.g., Transformer-based~\cite{xlsr,vaswani2017attention} and Conformer-based~\cite{Gulati2020-conformer}. The experimental setup  and results analysis (below)  for each proposed model refers to the results from Table~\ref{tab:asr-results}. \\

\textbf{Hybrid-based ASR:} for the hybrid-based ASR experiments, we use conventional biphone Convolutional Neural Network (CNN)~\cite{lecun1995convolutional} + TDNN-F~\cite{povey2018semi} based acoustic models trained with Kaldi~\cite{povey2011kaldi} toolkit (i.e., nnet3 model architecture) are used. AMs are trained with the LF-MMI training framework, considered to produce a state-of-the-art performance for hybrid ASR. In all the experiments, 3-fold speed perturbation with MFCCs and i-vectors features is used. LM is trained as a statistical 3-gram model using manual transcripts. Previous work related to ATC with this architecture is in~\cite{zuluaga2020automatic,zuluaga2022atco2}. \\

\textbf{XLSR-KALDI ASR:} in~\cite{9414741}, the authors propose to use LF-MMI criterion (similar to hybrid-based ASR) for the supervised adaptation of the self-supervised pretrained XLSR model~\cite{xlsr}. They also show that this approach outperforms the models trained with only the supervised data. Following that technique, we use the XLSR~\cite{babu2021xls} model pre-trained with a dataset as large as 50k\,hr of speech data, and later we fine-tune it with the supervised ATC data using the LF-MMI criterion. Further details about the architecture and experimental setup for pre-training the XLSR model can be found in the original paper~\cite{xlsr}. The results for this model are in the row tagged as '\texttt{XLSR-KALDI}' in Table~\ref{tab:asr-results}. \\

\textbf{End-to-End ASR:} we use the SpeechBrain~\cite{ravanelli2021speechbrain} toolkit to train a Conformer~\cite{Gulati2020-conformer} ASR model with ATC audio data. The Conformer model is composed of 12 encoder layers and additional 4 decoder layers (transformer-based~\cite{vaswani2017attention}).\footnote{We reuse the \texttt{Conformer-small} recipe from LibriSpeech~\cite{panayotov_librispeech_ICASSP2015} and adapted it to the ATC domain. See the recipe in: \url{https://github.com/speechbrain/speechbrain/tree/develop/recipes/LibriSpeech/ASR/transformer}} The dimension of the encoder and decoder model is set to $d_{model}=144$ with $d_{ffn}=d_{model}*4$. This accounts for a total of 11M parameters. We use dropout~\cite{srivastava2014dropout} with a probability of $dp=0.1$ for the attention and hidden layers, while Gaussian error linear units (GELU) is used as activation function~\cite{hendrycks2016gaussian}. We use Adam~\cite{kingma2014adam} optimizer with an initial learning rate $\gamma=1\mathrm{e}{-3}$.
We also use the default dynamic batching, which speeds up the training. During training, we combine per-frame conformer decoder output and CTC probabilities~\cite{karita2019comparative}. The CTC loss~\cite{ctc_loss} is weighted by $\alpha=0.3$. During inference and evaluation, the beam size is set to 66 with a CTC weight of $ctc_{w}=0.4$.

\begin{table}[t]
    \caption{WER for various \textcolor{Blue}{public} and \textcolor{Pink}{private} test sets with different ASR engines. The top results per block are \textbf{highlighted in bold}. The best result per test set is marked with \underline{underline}.
    $^{\ddagger}$ datasets from HAAWAII corpus and
    $^{\dagger}$datasets from MALORCA project~\cite{helmke2016reducing}.}
    \centering
    \label{tab:asr-results}
    \begin{tabular}{l | cccccc}
        \toprule
        \cellcolor{Gray}\textbf{Model} & \multicolumn{6}{c}{\cellcolor{Gray} \textbf{Test sets}} \\
        \cmidrule(lr){2-7}         
        \cellcolor{Gray} & \cellcolor{Pink}\textbf{NATS$^{\ddagger}$} & \cellcolor{Pink}\textbf{ISAVIA$^{\ddagger}$} & \cellcolor{Pink}\textbf{Prague$^{\dagger}$} & \cellcolor{Pink}\textbf{Vienna$^{\dagger}$} & \cellcolor{Blue}\textbf{ATCO2-1h }& \cellcolor{Blue}\textbf{ATCO2-4h} \\
        \midrule
        & \multicolumn{6}{l}{\cellcolor{Gray} \textbf{scenario~a) only supervised data}} \\
        \midrule
        CNN-TDNNF & 7.5 & 12.4 & 6.6 & 6.3 & 27.4 & 36.6 \\
        XLSR-KALDI & \underline{\textbf{7.1}} & \underline{\textbf{12.0}} & 6.7 & \underline{\textbf{5.5}} & \textbf{18.0} & \textbf{25.7} \\
        CONFORMER & 9.5 & 13.7  & \underline{\textbf{ 5.7}} & 7.0 & 41.8 & 46.2 \\
        \midrule
        &\multicolumn{6}{l}{\cellcolor{Gray} \textbf{ scenario~b) only ATCO2-PL 500\,hr data}} \\
        \midrule
        CNN-TDNNF & 26.7 & 34.1 & 11.7 & \textbf{11.8} & 19.1 & 25.1 \\
        CONFORMER & \textbf{21.6} & \textbf{32.5} &\textbf{ 7.6} & 12.5 & \underline{\textbf{15.9}} & \underline{\textbf{24.0}} \\
        \bottomrule
        \end{tabular}
\end{table}

\subsubsection{Training and Test Data}
\label{subsec:training_data}

\textbf{Training data configuration:} to see the effectiveness of using automatically transcribed data, as well as comparing performance on the in-domain VS out-of-domain sets, we trained both hybrid (CNN-TDNNF) and E2E (Conformer) models twice. First, we employ a mix between public and private supervised (recordings with gold annotations) ATC resources, which comprises around 190\,h.  We tag these models as, \textbf{\textit{scenario~a) only supervised data}}.
Second, we use a subset of 500\,hr of pseudo-annotated recordings (a seed ASR system is used to transcribe ATC communications from different airports) from the open-source ATCO2-PL set corpus (see introductory paper~\cite{zuluaga2022atco2}). We tag this model as,  \textit{\textbf{scenario~b) only ATCO2-PL 500\,h data}}. Results referencing both scenarios are in Table~\ref{tab:asr-results} and~\ref{tab:nats-boosting}. \\

\textbf{Test data configuration:} six different test sets are used for ASR evaluation, as shown in Table~\ref{tab:asr-results}. The first four test sets (highlighted in orange) are private and the last two test sets (highlighted in blue) are open data. Each two consecutive test sets are taken from one project: (i) NATS and ISAVIA are part of the HAAWAII corpus, (ii) Prague and Vienna are part of the MALORCA corpus and, (iii) ATCO2-1h and ATCO2-4h is from the ATCO2 project. 
Each dataset along with the test split is described in Section~\ref{sec:datasets}. We aimed at evaluating how the model's architecture, training paradigm (hybrid-based and E2E-ASR), and training data affect directly the performance of ASR. 

\subsubsection{Evaluation metric}

The preeminent technique for evaluating the efficacy of an ASR system is the word error rate (WER). This metric entails a meticulous comparison between the transcription of an utterance and the word sequence hypothesized by the ASR model. WER is determined by computing the aggregate of three types of errors, specifically, substitutions (S), insertions (I), and deletions (D), over the total count of words within the transcription. Should the reference transcript comprise N words, the WER can be computed using Equation~\ref{eq:wer}, outlined below.

\begin{equation}
    \label{eq:wer}
    WER=\frac{I + D + S}{N} \times 100.
\end{equation}

\noindent We evaluate all the models from Table~\ref{tab:asr-results} and Table~\ref{tab:nats-boosting} with WERs. For the boosting experiments (see Section~\ref{context_biasing} and Table~\ref{tab:nats-boosting}) we additionally use \textit{EntWER} which evaluates WER only on the callsign word sequence and \textit{ACC} which evaluates the accuracy of the system in capturing the target callsign in ICAO format.

\subsubsection{Speech Recognition Results}

The results of all the compared ASR models are in Table~\ref{tab:asr-results}.

\textbf{CNN-TDNNF model}: is our default architecture, as it has already been shown to work properly on the ATC domain. It has also been used largely in prior ATC work, like ATCO2, HAAWAII, and MALORCA (see Section~\ref{sec:introduction}). In our experiments, we trained this model for both \textit{scenario~a)} and \textit{scenario~b)}. Not surprisingly, we noted that the WERs are heavily impacted by the training and testing data. If we compare CNN-TDNNF  scenario~a) VS scenario~b), we see a systematical drop in performance for NATS 
\mbox{(7.5\% WER $\rightarrow$ 26.7\% WER)} and 
ISAVIA \mbox{(12.4\% WER $\rightarrow$ 34.1\% WER)}. However, in Prague and Vienna, which are still out-of-domain for scenario~b), less degradation in WERs is seen, 
\mbox{6.6\% WER $\rightarrow$ 11.7\% WER} for Prague and 
\mbox{6.3\% WER $\rightarrow$ 11.8\% WER} for Vienna. \\

\textbf{XLSR-KALDI model:} as mentioned earlier, we fine-tune the XLSR model (pre-trained based on wav2vec 2.0~\cite{wav2vec}) with the supervised data from the \textit{scenario~a)}. We do this as a proof of concept. The results show that the performance is consistent over all the private test sets compared to the CNN-TDNNF model trained with the same data. Though the model hasn't seen the noisy ATCO2 data during fine-tuning, since this model is pre-trained with large amounts of data, the WER on the ATCO2 test sets significantly improves compared to the CNN-TDNNF model. 
We see an absolute improvement of 9.4\% (27.4\% $\rightarrow$ 18\%) and 10.9\% (36.6\% $\rightarrow$ 25.7\%) for the ATCO2-1h and ATCO2-4h test set respectively. \\

\textbf{Conformer model:} we evaluate Conformer~\cite{Gulati2020-conformer}, an encoder-decoder Transformer-based~\cite{vaswani2017attention} model. With the Conformer architecture, we again train two models: on supervised data (\textit{scenario~a})) and on 500\,hr ATCO2-PL set (\textit{scenario~b)}). Both are tagged as \textit{CONFORMER} in Table~\ref{tab:asr-results}. Likewise, CNN-TDNNF models, the first four test sets are deemed in-domain for the baseline model, whereas the last two test sets (ATCO2-1h and ATCO2-4h test sets) are considered out-of-domain. Conversely, the second model is optimized for the out-of-domain test sets, while the first four are considered out-of-domain. Our goal is to demonstrate the effectiveness of the \textit{ATCO2-PL} dataset as an optimal resource for training models when only limited in-domain data is available. The second model demonstrates competitive performance when tested on close-mic speech datasets such as Prague and Vienna, which exclusively use the ATCo recordings. Yet, the model's performance deteriorates on more complex datasets, such as NATS and ISAVIA, which include pilot speech. We also note significant improvements on ATCO2-1h and ATCO2-4h test sets when training with the ATCO2-PL dataset. Scenario~b) exhibits a 62\% and 48\% relative WER reduction compared to scenario~a) on ATCO2-1h and ATCO2-4h, respectively. Whereas the first model performed poorly on both, 41.8 and 46.2\% WER, respectively. A critical consideration arises when examining the performance of the Conformer and CNN-TDNNF models under the same training scenario, \textit{scenario~b)}. Notably, the Conformer model outperforms the CNN-TDNNF model across all datasets, except for the Vienna test set. This leads us to hypothesize that the Conformer architecture shows greater proficiency when being trained over extensive datasets when compared to the CNN-TDNNF model in this particular scenario. \\

\begin{table}[t]
    \caption{Results for boosting on NATS test set corpus (HAAWAII). We ablate two models, scenario~a) a general ATC model trained only on supervised data, and scenario~b) a model trained on ATCO2-PL 500\,hr set. Results are obtained with offline CPU decoding. $^\mathparagraph$word error rates only on the sequence of words that compose the callsign in the utterance.}
    \centering
    \label{tab:nats-boosting}
    \resizebox{1\textwidth}{!}{
    \begin{tabular}{l | ccc| ccc}
        \toprule
        \rowcolor{Gray} \multicolumn{1}{c}{\textbf{Boosting}} & \multicolumn{3}{c}{\textbf{General ATC model}} & \multicolumn{3}{c}{\textbf{ATCO2 model-500h}} \\
        \cmidrule(lr){2-4} \cmidrule(lr){5-7}
         & WER & EntWER$^\mathparagraph$ & ACC & WER & EntWER$^\mathparagraph$ & ACC \\
        \midrule
        \rowcolor{Gray}\multicolumn{1}{c}{} & \multicolumn{3}{|c|}{\textbf{scenario~a) only supervised data}} & \multicolumn{3}{c}{\textbf{scenario~b) only ATCO2-PL 500\,hr data}}\\
        \midrule
        Baseline & 7.4 & 4.1 & 86.7 & 26.7 & 30.0 & 39.9 \\
        Unigrams & 7.4 & 3.6 & 88.0 & 25.6 & 24.1 & 46.2 \\
        N-grams & 6.7 & 2.0 & 93.3 & 23.4 & 19.8 & 61.3 \\
        GT boosted & 6.4 & 1.3 & 96.1 & 22.0 & 16.2 & 70.0 \\
        \bottomrule
    \end{tabular}
    }
\end{table}

\textbf{Callsign boosting with surveillance data ( $\approx$ contextual biasing)}: the contextual biasing approach is introduced in Section~\ref{context_biasing}. Table~\ref{tab:nats-boosting} demonstrates the effect of callsign boosting on the NATS test set (part of HAAWAII). The results of two ASR models are compared. Both models have the same architecture (Kaldi CNN-TDNNf) but are trained on different data. The first scenario~a), as in the experiments above, is trained on a combination of open-source and private annotated ATC databases which includes in-domain data (NATS); the second scenario~b) is trained on the 500 hours of automatically transcribed data collected and prepared for the ATCO2 project that is out-of-domain data. As expected the "in-domain" model performs better on the NATS dataset. At the same time, for both models, we can see a considerable improvement when contextual biasing is applied. The best results are achieved when only a ground truth callsign is boosted, i.e., 86.7\% $\rightarrow$ 96.1\% ACC for scenario~a) and 39.9\% $\rightarrow$ 70.0\% ACC for scenario~b). As in real life, we usually do not have the ground truth information, the improvement we can realistically get with radar data is shown in the line tagged as \underline{\textbf{N-grams}}. The effectiveness of biasing also depends on the number of callsigns used to build the biasing FST, as the more false callsigns are boosted the noisier the final rescoring is. According to previous findings, the ideal size of the biasing FST for improving performance depends on the data, but typically, performance begins to decline when there are more than 1000 contextual entities \cite{chen2019end}. In our data, we have an average of ~200 contextual entities per spoken phrase. For the n-grams boosting experiments, we achieved a relative improvement in Callsign WERs of 51.2\% and 34\%for callsigns recognition with models~a) and b), and 9.5\% and 12.4\% for the entire utterance, respectively (see Table~\ref{tab:nats-boosting}).

\subsection{High-level Entity Parser}
\label{subsec:results-ner}

A NER system is trained to parse text into high-level entities relevant to ATC communications.
The NER module (or tagger) is depicted in Figure~\ref{fig:metadata_pipeline}. First, a BERT\footnote{The pre-trained version of \texttt{BERT-base-uncased} with 110 million parameters is used. URL: \texttt{\url{https://huggingface.co/bert-base-uncased}}.}~\cite{devlin2018bert} model is downloaded from HuggingFace~\cite{wolf2020transformers,lhoest2021datasets} which is then fine-tuned on the NER task with 3k sentences ($\sim$3 hours of speech) using the \textit{ATCO2 test set corpus}. In this corpus, each word has a tag that corresponds to either:
\textcolor{teal}{\dashuline{callsign}}, 
\textcolor{red}{\dotuline{command}}, 
\textcolor{blue}{\uwave{values}}, and
\textcolor{black}{\uwave{UNK}} (everything else).
The final layer of the BERT model is replaced by a linear layer with a dimension of 8.\footnote{This setup follows the class structures from Section 3.3 of~\cite{zuluaga2021bertraffic}, two outputs for each class.} As only 3k sentences are used, a 5-fold cross-validation is conducted to avoid overfitting.
Further details about experimentation are covered in~\cite{nigmatulina2022two}. We redirect the reader to the public and open-source GitHub repository of the ATCO2 corpus.\footnote{ATCO2 GitHub repository: \url{https://github.com/idiap/atco2-corpus}} \\

\textbf{Experimental setup:} we fine-tune each model on an NVIDIA GeForce RTX 3090 for 10k steps. During experimentation, we use the same learning rate of $\gamma=5\mathrm{e}{-5}$ with a linear learning rate scheduler. Dropout~\cite{srivastava2014dropout} is set to $dp=0.1$ for the attention and hidden layers, while GELU is used as an activation function~\cite{hendrycks2016gaussian}. We also employ gradient norm clipping~\cite{pascanu2013difficulty}. We fine-tune each model with an effective batch size of 32 for 50 epochs with AdamW optimizer~\cite{loshchilov2018decoupled} ($\beta_1{=}0.9, \beta_2{=}0.999$, $\epsilon{=}1\mathrm{e}{-8}$). \\

\textbf{Evaluation metric:} we evaluate our NER system with F-score. The F-score or F-measure is a statistical measure utilized in binary classification analysis to evaluate a test's accuracy. The F1-score, defined in Equation~\ref{eq:f1-score}, represents the harmonic mean of precision and recall. Precision, as described in Equation~\ref{eq:precision}, is the ratio of true positive ($TP$) results to all positive results (including false positives ($FP$)), while recall, as defined in Equation~\ref{eq:recall}, is the ratio of $TP$ to all samples that should have been identified as positive (including false negatives ($FN$)):

\begin{equation}
    \label{eq:precision}
    Precision = \frac{TP}{TP+FP}
\end{equation}

\begin{equation}
    \label{eq:recall}
    Recall = \frac{TP}{TP+FN}
\end{equation}

\begin{equation}
    \label{eq:f1-score}
    F1 = \frac{2*Precision*Recall}{Precision+Recall} = \frac{2*TP}{2*TP+FP+FN}
\end{equation}

\vspace{0.2cm}

\textbf{Results:} the \textit{\textbf{high-level entity parser}} system is evaluated on the only available public resource, the ATCO2-4h test set, which contains word level tags, i.e., callsign, command, and values. The results for precision, recall, and F1-score, over each of the proposed classes are listed in Table~\ref{tab:ner_results}. Our BERT-based system achieves a high level of performance in callsign detection, with an F1-Score of 97.1\%. However, the command and values classes show an average worse performance, with F1-scores of 82.0\% and 87.2\%, respectively. Notably, the command class presents the greatest challenge due to its inherent complexity when compared to values and callsigns. Values are predominantly composed of keywords, such as “flight level” followed by cardinal numbers such as “two”, “three hundred”, or “one thousand”. These characteristics make them easy for a system to recognize. Similarly, callsigns are highly structured, consisting of an airline designator accompanied by numbers and letters spoken in radiotelephony alphabet~\cite{allclear}. Given their importance in communication, as in callsign highlighting~\cite{zuluaga2022atco2} or read-back error detection~\cite{hartmut2022readback}, additional validation is necessary for real-life scenarios, or when working with proprietary/private data.

Although our BERT-based system achieves high performance in callsign recognition, there is still room for improvement. One potential method for enhancing performance is to incorporate real-time surveillance data into the system~\cite{nigmatulina2022two}, which was introduced in Section~\ref{context_biasing} and Table~\ref{tab:nats-boosting}. \\

\begin{table}[t]
    \caption{F1-score-@F1, Precision-@P, and Recall-@R metrics for callsign, command, and values classes of the \textit{\textbf{high-level entity parser}} system. Results are averaged over a 5-fold cross-validation scheme on \textit{ATCO2-4h corpus} in order to mitigate overfitting. We run five times fine-tuning with different training seeds (2222/3333/4444/5555/6666). }
    \centering
    \label{tab:ner_results}
    \begin{tabular}{l ccc ccc ccc}
        \toprule
        Model & \multicolumn{3}{c}{\textbf{\textcolor{teal}{Callsign}}} & \multicolumn{3}{c}{\textbf{\textcolor{purple}{Command}}} & \multicolumn{3}{c}{\textbf{\textcolor{red}{Values}}} \\
        \cmidrule(lr){2-4}
        \cmidrule(lr){5-7}
        \cmidrule(lr){8-10}
        & @P & @R & @F1 & @P & @R & @F1 & @P & @R & @F1 \\
        \midrule
        \texttt{bert-base-uncased} & 97.1 & 97.8 & 97.5 & 80.4 & 83.6 & 82.0 & 86.3 & 88.1 & 87.2 \\
        \bottomrule
    \end{tabular}
\end{table}

\subsection{Response Generator}

The response generator is composed of 3 submodules, see Section~\ref{subsec:response-generator}. The \textbf{Grammar Conversion} and \textbf{Word Fixer} submodules are based on hard-coded rules. Thus, we do not provide any quantitative results. 

\subsection{Text to Speech}

Text-to-speech (TTS) is a technology that facilitates the conversion of written text into spoken language. When employed in ATC, TTS can be integrated with virtual simulation-pilot systems to train ATCos. In our study, we utilize a state-of-the-art non-autoregressive speech synthesizer, the FastSpeech2 model~\cite{ren2020fastspeech}. To accomplish this, we download and utilize a widely recognized pre-trained TTS model from the HuggingFace hub~\cite{wolf2020transformers}.\footnote{Access to the model can be obtained at \url{https://huggingface.co/facebook/fastspeech2-en-ljspeech}} FastSpeech2 is used in inference mode with the sentence produced by the grammar conversion submodule, and subsequently, the word fixer submodule, serving as the prompt of the virtual simulation-pilot. Other models, such as Tacotron~\cite{Wang2017tacotron} or Tacotron2~\cite{shen2018tacotron2},\footnote{Free access to the Tacotron2 model can be found at \texttt{\url{https://huggingface.co/speechbrain/tts-tacotron2-ljspeech}}.} can be fine-tuned and implemented to handle ATC data. \\

\textbf{System analysis:} in our experiments, we discovered that the model is capable of handling complex word sequences, such as those commonly encountered in ATC, including read backs from virtual simulation-pilots that contain multiple commands and values. However, we did not conduct any qualitative analysis of the TTS-produced voice or speech, leaving this as a future area of exploration. We also did not investigate the possibility of fine-tuning the TTS module with ATC audio data, as our main focus was on developing a simple and effective virtual simulation-pilot system using pre-existing open-source models. \\

\textbf{Future line of work:} although we did not pursue this area, it is indeed possible to fine-tune the TTS module using in-domain ATC data. In Table~\ref{tab:databases}, we provide a list of both public and private databases that could be utilized for this purpose. Generally, the same annotations used for ASR can also be applied to fine-tune a TTS system. However, there are two different approaches that could be explored simultaneously or sequentially. First, by considering the speaker role (ATCo or pilot), the TTS module could be biased to produce speech that is more appropriate for different roles, such as noisy speech from pilots. Second, if datasets are available that provide information on gender and accents, such as the ATCOSIM dataset~\cite{ATCOSIM}, TTS models with different accents and gender could be developed.

\section{Limitations \& Future Work}
\label{sec:future-work}

In our investigation of ASR systems, we have explored the potential of hybrid-based and E2E ASR systems, which can be further enhanced by incorporating data relevant to the specific exercise undertaken by ATCo trainees, such as runway numbers or waypoint lists. Moving forward, we suggest that research should continue to explore E2E training techniques for ASR, as well as methods for integrating contextual data into these E2E systems.

The repetition generator currently in use employs a simple grammar converter and a pre-trained TTS system. However, we believe that additional efforts could be made to enhance the system's ability to convey more complex ATC communications to virtual simulation-pilots. In particular, the TTS system could be fine-tuned to produce female or male voices, as well as modify key features such as speech rate, noise artifact, or cues to synthesize voice in a stressful situation. Additionally, a quantitative metric for evaluating the TTS system could be integrated to further enhance its efficacy. We also list some optional modules (see Section~\ref{subsec:optional-mod}) which can be further explored, e.g., the read back insertion module or voice activity detection. 

Similarly, there is scope for the development of multimodal and multitask systems. Such systems would be fed with real-time ATC communications and contextual data simultaneously, later generating transcripts and high-level entities as the output. Such systems could be considered a “dual ASR and high-level entity parser”. Finally, legal and ethical challenges of using ATC audio data is another important field that needs to be further explored in future work. We redirect the reader to the \textbf{Legal and privacy aspects for collection of ATC recordings} section in~\cite{zuluaga2022atco2} and to previous work in~\cite{rigault2022legal}.

\section{How to Develop your Own Virtual Simulation-Pilot?}

If you would like to replicate this work with in-domain data, i.e., for a specific scenario or airport, you can follow the steps below:

\begin{enumerate}
    \item Start by defining the set of rules and grammar to use for the annotation protocol. You can follow the cheat-sheet from ATCO2 project~\cite{zuluaga2022atco2}. See \url{https://www.spokendata.com/atc} and \url{https://www.atco2.org/}. In addition, one can use previous ontologies developed for ATC~\cite{helmke2018ontology}.
    \item For training or adapting the virtual simulation-pilot engine, you need three sets of annotations: (i)~gold annotations of the ATCo-pilot communications for ASR adaptation; (ii)~high-level entity annotations (callsign, command and values) to perform NLU for ATC; and (iii)~a set of rules to convert ATCo commands into 'pilots' read backs, e.g., \textcolor{blue}{"\dashuline{descend to}"} $\rightarrow$ \textcolor{red}{"\dashuline{descending to}"}.
    \item Gather and annotate at least 1 hour of ATCo speech, and 1k\,samples for training your \textit{\textbf{high-level entity parser}} system. If the reader is interested in getting a general idea of how much data is needed for reaching a desired WER or F1-score, you can check~\cite{zuluaga2022does} for ASR and~\cite{zuluaga2022atco2} for ATC-NLU. 
    \item Fine-tune a strong pre-trained ASR model, e.g., Wav2Vec 2.0 or XLSR~\cite{wav2vec,xlsr} with the ATC audio recordings.\footnote{If the performance is not sufficient, you can use open-source corpora (see Table~\ref{tab:databases}) to increase the amount of annotated samples, see~\cite{LDC_ATCC,zuluaga2022atco2,UWB_ATCC,ATCOSIM}.} We recommend acquiring the ATCO2-PL dataset~\cite{zuluaga2022atco2}, which has proven to be a good starting point when no in-domain data is available. This is related to ASR and NLU for ATC. 
    \item Fine-tune a strong pre-trained NLP model, e.g., BERT~\cite{devlin2018bert} or RoBERTa~\cite{liu2019roberta} with the NLP tags. If the performance is not sufficient, you can follow several data augmentation techniques on the text level. For example, you can replace the callsign in one training sample by different ones from a predefined callsign list. In that way, one can generate many more valuable training samples. It is also possible to use more annotations during fine-tuning, e.g., see \texttt{ATCO2-4h} corpus~\cite{zuluaga2022atco2}.
    \item Lastly, in case you need to adapt the TTS module to pilot speech, you could adapt the FastSpeech2~\cite{ren2020fastspeech} system. Then, you need to invert the annotations used for ASR, i.e., using the transcripts as input and the ATCo or pilot recordings as 'targets'. This step is not strictly necessary, as already available pre-trained modules possess a good quality. 
\end{enumerate}

\section{Conclusion}
\label{sec:conclusion}

In this paper, we have presented a novel virtual simulation-pilot system designed for ATCo training. Our system utilizes cutting-edge open-source ASR, NLP, and TTS systems. To the best of our knowledge, this is the first such system that relies on open-source ATC resources. The virtual simulation-pilot system is developed for ATCo training purposes, thus, this work represents an important contribution to the field of aviation training.

Our system employs a multi-stage approach, including ASR transcription, \textit{\textbf{high-level entity parser}} system, and a repetition generator module to provide pilot-like responses to ATC communications. By utilizing open-source AI models and public databases, we have developed a simple and efficient system that can be easily replicated and adapted for different training scenarios. For instance, we tested our ASR system on different well-known ATC-related projects i.e., HAAWAII, MALORCA and ATCO2. We reached as low as 5.5\% WERs on high-quality data (MALORCA, ATCo speech in operations room) and 15.9\% WERs on low-quality ATC audio such as the test sets from ATCO2 project (noise levels below 15 dB).

Going forward, there is significant potential for further improvements and expansions to our system. Incorporating contextual data, such as runway numbers or waypoint lists, could enhance the accuracy and effectiveness of the ASR and high-level entity parser modules. We performed evaluations with  real-time surveillance, which proved to further improve the system's performance in recognizing and responding to ATC communications. For instance, our boosting technique brings a 9\% absolute melioration in callsign detection accuracy levels (86.7\% $\rightarrow$ 96.1\%) for NATS test set. 

We also believe that additional efforts can be made to fine-tune the TTS system for improved synthesis of male or female voices, as well as modifying speech rate, noise artifact, and other features.

Our ASR system can reach as low as 5.5\% and 15.9\% word error rates (WER) on high and low-quality ATC audio (Vienna and ATCO2-test-set-1h, respectively). We also demonstrate that adding surveillance data into the ASR can yield callsign detection accuracy of more than 96\%. Overall, this work represents a promising first step towards developing advanced virtual simulation-pilot systems for ATCo training, and we expect future work will continue to explore this field.


\section*{Author Contributions}

Conceptualization, Juan Pablo Zuluaga Gomez, Amrutha Prasad, Iuliia Nigmatulina and Petr Motlicek; Data curation, Juan Pablo Zuluaga Gomez, Amrutha Prasad and Iuliia Nigmatulina; Formal analysis, Juan Pablo Zuluaga Gomez, Iuliia Nigmatulina and Matthias Kleinert; Funding acquisition, Petr Motlicek; Investigation, Juan Pablo Zuluaga Gomez, Amrutha Prasad and Iuliia Nigmatulina; Methodology, Juan Pablo Zuluaga Gomez, Amrutha Prasad, Iuliia Nigmatulina, Petr Motlicek and Matthias Kleinert; Project administration, Petr Motlicek; Resources, Juan Pablo Zuluaga Gomez and Amrutha Prasad; Software, Juan Pablo Zuluaga Gomez, Amrutha Prasad, Iuliia Nigmatulina and Petr Motlicek; Supervision, Petr Motlicek; Validation, Juan Pablo Zuluaga Gomez and Iuliia Nigmatulina; Visualization, Juan Pablo Zuluaga Gomez; Writing – original draft, Juan Pablo Zuluaga Gomez, Iuliia Nigmatulina and Matthias Kleinert; Writing – review \& editing, Juan Pablo Zuluaga Gomez, Petr Motlicek and Matthias Kleinert.

\section{Bibliography}
\bibliography{biblio}

\begin{thebibliography}{999}

\bibitem[Nassif \em{et~al.}(2019)Nassif, Shahin, Attili, Azzeh, and
  Shaalan]{nassif2019speech}
Nassif, A.B.; Shahin, I.; Attili, I.; Azzeh, M.; Shaalan, K.
\newblock Speech recognition using deep neural networks: A systematic review.
\newblock {\em IEEE access} {\bf 2019}, {\em 7},~19143--19165.

\bibitem[Otter \em{et~al.}(2020)Otter, Medina, and Kalita]{otter2020survey}
Otter, D.W.; Medina, J.R.; Kalita, J.K.
\newblock A survey of the usages of deep learning for natural language
  processing.
\newblock {\em IEEE transactions on neural networks and learning systems} {\bf
  2020}, {\em 32},~604--624.

\bibitem[Zhang \em{et~al.}(2018)Zhang, Wang, and Liu]{zhang2018deep}
Zhang, L.; Wang, S.; Liu, B.
\newblock Deep learning for sentiment analysis: A survey.
\newblock {\em Wiley Interdisciplinary Reviews: Data Mining and Knowledge
  Discovery} {\bf 2018}, {\em 8},~e1253.

\bibitem[Lugosch \em{et~al.}(2019)Lugosch, Ravanelli, Ignoto, Tomar, and
  Bengio]{Lugosch2019}
Lugosch, L.; Ravanelli, M.; Ignoto, P.; Tomar, V.S.; Bengio, Y.
\newblock {Speech Model Pre-Training for End-to-End Spoken Language
  Understanding}.
\newblock In Proceedings of the Proc. Interspeech 2019,  2019, pp. 814--818.
\newblock {\url{https://doi.org/10.21437/Interspeech.2019-2396}}.

\bibitem[Beek \em{et~al.}(1977)Beek, Neuberg, and Hodge]{beek1977}
Beek, B.; Neuberg, E.; Hodge, D.
\newblock An assessment of the technology of automatic speech recognition for
  military applications.
\newblock {\em IEEE Transactions on Acoustics, Speech, and Signal Processing}
  {\bf 1977}, {\em 25},~310--322.

\bibitem[Hamel \em{et~al.}(1989)Hamel, Kotick, and Layton]{hamel1989}
Hamel, C.J.; Kotick, D.; Layton, M.
\newblock Microcomputer system integration for air control training.
\newblock Technical report, Naval Training Systems Center, Orlando FL,  1989.

\bibitem[Matrouf \em{et~al.}(1990)Matrouf, Gauvain, Neel, and
  Mariani]{matrouf1990adapting}
Matrouf, K.; Gauvain, J.; Neel, F.; Mariani, J.
\newblock Adapting probability-transitions in {DP} matching processing for an
  oral task-oriented dialogue.
\newblock In Proceedings of the International Conference on Acoustics, Speech,
  and Signal Processing (ICASSP). IEEE,  1990, pp. 569--572.

\bibitem[Weinstein(1991)]{weinstein1991opportunities}
Weinstein, C.J.
\newblock Opportunities for advanced speech processing in military
  computer-based systems.
\newblock {\em Proceedings of the IEEE} {\bf 1991}, {\em 79},~1626--1641.

\bibitem[Kleinert \em{et~al.}(2018)Kleinert, Helmke, Siol, Ehr, Cerna, Kern,
  Klakow, Motlicek, Oualil, Singh, et~al.]{kleinert2018semi}
Kleinert, M.; Helmke, H.; Siol, G.; Ehr, H.; Cerna, A.; Kern, C.; Klakow, D.;
  Motlicek, P.; Oualil, Y.; Singh, M.;  et~al.
\newblock Semi-supervised adaptation of assistant based speech recognition
  models for different approach areas.
\newblock In Proceedings of the 37th Digital Avionics Systems Conference
  (DASC). IEEE,  2018, pp. 1--10.

\bibitem[Kleinert \em{et~al.}(2019)Kleinert, Helmke, Siol, Ehr, Klakow, Singh,
  Motlicek, Kern, Cerna, and Hlousek]{kleinert2019adaptation}
Kleinert, M.; Helmke, H.; Siol, G.; Ehr, H.; Klakow, D.; Singh, M.; Motlicek,
  P.; Kern, C.; Cerna, A.; Hlousek, P.
\newblock Adaptation of assistant based speech recognition to new domains and
  its acceptance by air traffic controllers.
\newblock In Proceedings of the International Conference on Intelligent Human
  Systems Integration. Springer,  2019, pp. 820--826.

\bibitem[Helmke \em{et~al.}(2016)Helmke, Ohneiser, M{\"u}hlhausen, and
  Wies]{helmke2016reducing}
Helmke, H.; Ohneiser, O.; M{\"u}hlhausen, T.; Wies, M.
\newblock Reducing controller workload with automatic speech recognition.
\newblock In Proceedings of the 2016 IEEE/AIAA 35th Digital Avionics Systems
  Conference (DASC). IEEE,  2016, pp. 1--10.

\bibitem[Helmke \em{et~al.}(2017)Helmke, Ohneiser, Buxbaum, and
  Kern]{helmke2017increasing}
Helmke, H.; Ohneiser, O.; Buxbaum, J.; Kern, C.
\newblock Increasing {ATM} efficiency with assistant based speech recognition.
\newblock In Proceedings of the Proc. of the 13th USA/Europe Air Traffic
  Management Research and Development Seminar, Seattle, USA,  2017.

\bibitem[Kleinert \em{et~al.}(2022)Kleinert, Shetty, Helmke, Ohneiser, Wiese,
  Maier, Schacht, Nigmatulina, Saeed, and Sarfjoo]{kleinert2021apron}
Kleinert, M.; Shetty, S.; Helmke, H.; Ohneiser, O.; Wiese, H.; Maier, M.;
  Schacht, S.; Nigmatulina, I.; Saeed, S.; Sarfjoo, P.M.
\newblock Apron Controller Support by Integration of Automatic Speech
  Recognition with an Advanced Surface Movement Guidance and Control System
  {\bf 2022}.

\bibitem[Nigmatulina \em{et~al.}(2022)Nigmatulina, Zuluaga-Gomez, Prasad,
  Sarfjoo, and Motlicek]{nigmatulina2022two}
Nigmatulina, I.; Zuluaga-Gomez, J.; Prasad, A.; Sarfjoo, S.S.; Motlicek, P.
\newblock A two-step approach to leverage contextual data: speech recognition
  in air-traffic communications.
\newblock In Proceedings of the ICASSP,  2022.

\bibitem[Kleinert \em{et~al.}(2021)Kleinert, Helmke, Shetty, Ohneiser, Ehr,
  Prasad, Motlicek, and Harfmann]{kleinert2021automated}
Kleinert, M.; Helmke, H.; Shetty, S.; Ohneiser, O.; Ehr, H.; Prasad, A.;
  Motlicek, P.; Harfmann, J.
\newblock Automated Interpretation of Air Traffic Control Communication: The
  Journey from Spoken Words to a Deeper Understanding of the Meaning.
\newblock In Proceedings of the 2021 IEEE/AIAA 40th Digital Avionics Systems
  Conference (DASC). IEEE,  2021, pp. 1--9.

\bibitem[Zuluaga-Gomez \em{et~al.}(2022)Zuluaga-Gomez, Vesel{\`y}, Sz{\"o}ke,
  Motlicek, Kocour, Rigault, Choukri, Prasad, Sarfjoo, Nigmatulina,
  et~al.]{zuluaga2022atco2}
Zuluaga-Gomez, J.; Vesel{\`y}, K.; Sz{\"o}ke, I.; Motlicek, P.; Kocour, M.;
  Rigault, M.; Choukri, K.; Prasad, A.; Sarfjoo, S.S.; Nigmatulina, I.;  et~al.
\newblock ATCO2 corpus: A Large-Scale Dataset for Research on Automatic Speech
  Recognition and Natural Language Understanding of Air Traffic Control
  Communications.
\newblock {\em arXiv preprint arXiv:2211.04054} {\bf 2022}.

\bibitem[Kocour \em{et~al.}(2021{\natexlab{a}})Kocour, Veselý, Sz{\"o}ke,
  Kesiraju, Zuluaga-Gomez, Blatt, Prasad, Nigmatulina, Motl{\'\i}{\v{c}}ek,
  Klakow, et~al.]{kocour2021automatic}
Kocour, M.; Veselý, K.; Sz{\"o}ke, I.; Kesiraju, S.; Zuluaga-Gomez, J.; Blatt,
  A.; Prasad, A.; Nigmatulina, I.; Motl{\'\i}{\v{c}}ek, P.; Klakow, D.;  et~al.
\newblock Automatic processing pipeline for collecting and annotating
  air-traffic voice communication data.
\newblock {\em Engineering Proceedings} {\bf 2021}, {\em 13},~8.

\bibitem[Kocour \em{et~al.}(2021{\natexlab{b}})Kocour, Veselý, Blatt, Gomez,
  Sz{\"o}ke, Cernocky, Klakow, and Motlicek]{kocour21_interspeech}
Kocour, M.; Veselý, K.; Blatt, A.; Gomez, J.Z.; Sz{\"o}ke, I.; Cernocky, J.;
  Klakow, D.; Motlicek, P.
\newblock {Boosting of Contextual Information in ASR for Air-Traffic Call-Sign
  Recognition}.
\newblock In Proceedings of the Interspeech,  2021, pp. 3301--3305.
\newblock {\url{https://doi.org/10.21437/Interspeech.2021-1619}}.

\bibitem[Zuluaga-Gomez \em{et~al.}(2023)Zuluaga-Gomez, Sarfjoo, Prasad,
  Nigmatulina, Motlicek, Ondre, Ohneiser, and Helmke]{zuluaga2021bertraffic}
Zuluaga-Gomez, J.; Sarfjoo, S.S.; Prasad, A.; Nigmatulina, I.; Motlicek, P.;
  Ondre, K.; Ohneiser, O.; Helmke, H.
\newblock BERTraffic: BERT-based Joint Speaker Role and Speaker Change
  Detection for Air Traffic Control Communications.
\newblock {\em IEEE Spoken Language Technology Workshop (SLT), Doha, Qatar}
  {\bf 2023}.

\bibitem[Lin \em{et~al.}(2021)Lin, Li, Yang, Yan, Tan, and
  Chen]{lin2021improving}
Lin, Y.; Li, Q.; Yang, B.; Yan, Z.; Tan, H.; Chen, Z.
\newblock Improving speech recognition models with small samples for air
  traffic control systems.
\newblock {\em Neurocomputing} {\bf 2021}, {\em 445},~287--297.

\bibitem[Zuluaga-Gomez \em{et~al.}(2020)Zuluaga-Gomez, Vesel{\`y}, Blatt,
  Motlicek, Klakow, Tart, Sz{\"o}ke, Prasad, Sarfjoo, Kol{\v{c}}{\'a}rek,
  et~al.]{zuluaga2020automatic}
Zuluaga-Gomez, J.; Vesel{\`y}, K.; Blatt, A.; Motlicek, P.; Klakow, D.; Tart,
  A.; Sz{\"o}ke, I.; Prasad, A.; Sarfjoo, S.; Kol{\v{c}}{\'a}rek, P.;  et~al.
\newblock Automatic call sign detection: Matching air surveillance data with
  air traffic spoken communications.
\newblock In Proceedings of the Multidisciplinary Digital Publishing Institute
  Proceedings,  2020, Vol.~59, p.~14.

\bibitem[De~Cordoba \em{et~al.}(2006)De~Cordoba, Ferreiros, San-Segundo,
  Macias-Guarasa, Montero, Fernandez, D'Haro, and Pardo]{de2006air}
De~Cordoba, R.; Ferreiros, J.; San-Segundo, R.; Macias-Guarasa, J.; Montero,
  J.; Fernandez, F.; D'Haro, L.; Pardo, J.
\newblock Air traffic control speech recognition system cross-task \& speaker
  adaptation.
\newblock {\em IEEE Aerospace and Electronic Systems Magazine} {\bf 2006}, {\em
  21},~12--17.

\bibitem[Fan \em{et~al.}(2021)Fan, Guo, Lin, Yang, and Zhang]{fan2021speech}
Fan, P.; Guo, D.; Lin, Y.; Yang, B.; Zhang, J.
\newblock Speech recognition for air traffic control via feature learning and
  end-to-end training.
\newblock {\em arXiv preprint arXiv:2111.02654} {\bf 2021}.

\bibitem[Lopez \em{et~al.}(2013)Lopez, Condamines, Josselin-Leray, O'Donoghue,
  and Salmon]{enac}
Lopez, S.; Condamines, A.; Josselin-Leray, A.; O'Donoghue, M.; Salmon, R.
\newblock Linguistic analysis of English phraseology and plain language in
  air-ground communication.
\newblock {\em Journal of Air Transport Studies} {\bf 2013}, {\em 4},~44--60.

\bibitem[Graglia \em{et~al.}(2005)Graglia, Favennec, and Arnoux]{vocalise}
Graglia, L.; Favennec, B.; Arnoux, A.
\newblock {Vocalise: Assessing the impact of data link technology on the R/T
  channel}.
\newblock In Proceedings of the 24th Digital Avionics Systems Conference. IEEE,
   2005, Vol.~1, pp. 5--C.

\bibitem[Helmke \em{et~al.}(2018)Helmke, Slotty, Poiger, Herrer, Ohneiser,
  Vink, Cerna, Hartikainen, Josefsson, Langr, et~al.]{helmke2018ontology}
Helmke, H.; Slotty, M.; Poiger, M.; Herrer, D.F.; Ohneiser, O.; Vink, N.;
  Cerna, A.; Hartikainen, P.; Josefsson, B.; Langr, D.;  et~al.
\newblock {Ontology for transcription of ATC speech commands of SESAR 2020
  solution PJ. 16-04}.
\newblock In Proceedings of the IEEE/AIAA 37th Digital Avionics Systems
  Conference (DASC). IEEE,  2018, pp. 1--10.

\bibitem[\mbox{International Civil Aviation Organization}(2020)]{allclear}
\mbox{International Civil Aviation Organization}.
\newblock {ICAO} Phraseology Reference Guide,  2020.

\bibitem[Bouchal \em{et~al.}(2022)Bouchal, Had, and
  Bouchaudon]{bouchal2022design}
Bouchal, A.; Had, P.; Bouchaudon, P.
\newblock The Design and Implementation of Upgraded ESCAPE Light ATC Simulator
  Platform at the CTU in Prague.
\newblock In Proceedings of the 2022 New Trends in Civil Aviation (NTCA). IEEE,
   2022, pp. 103--108.

\bibitem[Lin(2021)]{lin2021spoken}
Lin, Y.
\newblock Spoken instruction understanding in air traffic control: Challenge,
  technique, and application.
\newblock {\em Aerospace} {\bf 2021}, {\em 8},~65.

\bibitem[Lin \em{et~al.}(2021)Lin, Wu, Guo, Zhang, Yin, Yang, and
  Zhang]{lin2021deep}
Lin, Y.; Wu, Y.; Guo, D.; Zhang, P.; Yin, C.; Yang, B.; Zhang, J.
\newblock A deep learning framework of autonomous pilot agent for air traffic
  controller training.
\newblock {\em IEEE Transactions on Human-Machine Systems} {\bf 2021}, {\em
  51},~442--450.

\bibitem[Prasad \em{et~al.}(2022)Prasad, Zuluaga-Gomez,
  et~al.]{prasad2022pseudo_pilot}
Prasad, A.; Zuluaga-Gomez, J.;  et~al.
\newblock {Speech and Natural Language Processing Technologies for Pseudo-Pilot
  Simulator}.
\newblock In Proceedings of the 12th SESAR Innovation Days. Sesar Joint
  Undertaking.,  2022.

\bibitem[Baevski \em{et~al.}(2020)Baevski, Zhou, Mohamed, and Auli]{wav2vec}
Baevski, A.; Zhou, Y.; Mohamed, A.; Auli, M.
\newblock wav2vec 2.0: {A} Framework for Self-Supervised Learning of Speech
  Representations.
\newblock In Proceedings of the Advances in Neural Information Processing
  Systems, December 6-12, 2020, virtual,  2020.

\bibitem[Babu \em{et~al.}(2021)Babu, Wang, Tjandra, Lakhotia, Xu, Goyal, Singh,
  von Platen, Saraf, Pino, et~al.]{xlsr}
Babu, A.; Wang, C.; Tjandra, A.; Lakhotia, K.; Xu, Q.; Goyal, N.; Singh, K.;
  von Platen, P.; Saraf, Y.; Pino, J.;  et~al.
\newblock Xls-r: Self-supervised cross-lingual speech representation learning
  at scale.
\newblock {\em ArXiv preprint} {\bf 2021}, {\em abs/2111.09296}.

\bibitem[Devlin \em{et~al.}(2018)Devlin, Chang, Lee, and
  Toutanova]{devlin2018bert}
Devlin, J.; Chang, M.W.; Lee, K.; Toutanova, K.
\newblock {BERT: Pre-training of Deep Bidirectional Transformers for Language
  Understanding}.
\newblock {\em arXiv preprint arXiv:1810.04805} {\bf 2018}.

\bibitem[Ren \em{et~al.}(2021)Ren, Hu, Tan, Qin, Zhao, Zhao, and
  Liu]{ren2020fastspeech}
Ren, Y.; Hu, C.; Tan, X.; Qin, T.; Zhao, S.; Zhao, Z.; Liu, T.
\newblock FastSpeech 2: Fast and High-Quality End-to-End Text to Speech.
\newblock In Proceedings of the 9th International Conference on Learning
  Representations, {ICLR} 2021, Virtual Event, Austria, May 3-7, 2021.
  OpenReview.net,  2021.

\bibitem[Godfrey(1994)]{LDC_ATCC}
Godfrey, J.
\newblock {The Air Traffic Control Corpus (ATC0) - LDC94S14A},  1994.

\bibitem[{\v{S}}m{\'\i}dl \em{et~al.}(2019){\v{S}}m{\'\i}dl, {\v{S}}vec,
  Tihelka, Matou{\v{s}}ek, Romportl, and Ircing]{UWB_ATCC}
{\v{S}}m{\'\i}dl, L.; {\v{S}}vec, J.; Tihelka, D.; Matou{\v{s}}ek, J.;
  Romportl, J.; Ircing, P.
\newblock Air traffic control communication ({ATCC}) speech corpora and their
  use for {ASR} and {TTS} development.
\newblock {\em Language Resources and Evaluation} {\bf 2019}, {\em
  53},~449--464.

\bibitem[Hofbauer \em{et~al.}(2008)Hofbauer, Petrik, and Hering]{ATCOSIM}
Hofbauer, K.; Petrik, S.; Hering, H.
\newblock The {ATCOSIM} Corpus of Non-Prompted Clean Air Traffic Control
  Speech.
\newblock In Proceedings of the LREC,  2008.

\bibitem[Pavlinovi{\'c} \em{et~al.}(2017)Pavlinovi{\'c}, Juri{\v{c}}i{\'c}, and
  Antulov-Fantulin]{pavlinovic2017air}
Pavlinovi{\'c}, M.; Juri{\v{c}}i{\'c}, B.; Antulov-Fantulin, B.
\newblock Air traffic controllers' practical part of basic training on computer
  based simulation device.
\newblock In Proceedings of the International Convention on Information and
  Communication Technology, Electronics and Microelectronics (MIPRO). IEEE,
  2017, pp. 920--925.

\bibitem[Juri{\v{c}}i{\'c} \em{et~al.}(2011)Juri{\v{c}}i{\'c}, Vare{\v{s}}ak,
  and Bo{\v{z}}i{\'c}]{jurivcic2011role}
Juri{\v{c}}i{\'c}, B.; Vare{\v{s}}ak, I.; Bo{\v{z}}i{\'c}, D.
\newblock The role of the simulation devices in air traffic controller
  training.
\newblock In Proceedings of the International Symposium on Electronics in
  Traffic, ISEP 2011 Proceedings,  2011.

\bibitem[Chhaya \em{et~al.}(2018)Chhaya, Jafer, Coyne, Thigpen, and
  Durak]{chhaya2018enhancing}
Chhaya, B.; Jafer, S.; Coyne, W.B.; Thigpen, N.C.; Durak, U.
\newblock Enhancing scenario-centric air traffic control training.
\newblock In Proceedings of the 2018 AIAA modeling and simulation technologies
  conference,  2018, p. 1399.

\bibitem[Updegrove and Jafer(2017)]{updegrove2017optimization}
Updegrove, J.A.; Jafer, S.
\newblock Optimization of air traffic control training at the Federal Aviation
  Administration Academy.
\newblock {\em Aerospace} {\bf 2017}, {\em 4},~50.

\bibitem[Eide \em{et~al.}(2014)Eide, {\O}deg{\aa}rd, and
  Karahasanovi{\'c}]{eide2014post}
Eide, A.W.; {\O}deg{\aa}rd, S.S.; Karahasanovi{\'c}, A.
\newblock A post-simulation assessment tool for training of air traffic
  controllers.
\newblock In Proceedings of the International Conference on Human Interface and
  the Management of Information. Springer,  2014, pp. 34--43.

\bibitem[N{\'e}methov{\'a} \em{et~al.}(2019)N{\'e}methov{\'a}, B{\'a}lint, and
  Vagner]{nemethova2019education}
N{\'e}methov{\'a}, H.; B{\'a}lint, J.; Vagner, J.
\newblock The education and training methodology of the air traffic controllers
  in training.
\newblock In Proceedings of the International Conference on Emerging eLearning
  Technologies and Applications (ICETA). IEEE,  2019, pp. 556--563.

\bibitem[Zhang \em{et~al.}(2022)Zhang, Zhang, Guo, Zhou, Wu, Yang, and
  Lin]{zhang2022automatic}
Zhang, J.; Zhang, P.; Guo, D.; Zhou, Y.; Wu, Y.; Yang, B.; Lin, Y.
\newblock Automatic repetition instruction generation for air traffic control
  training using multi-task learning with an improved copy network.
\newblock {\em Knowledge-Based Systems} {\bf 2022}, {\em 241},~108232.

\bibitem[Vaswani \em{et~al.}(2017)Vaswani, Shazeer, Parmar, Uszkoreit, Jones,
  Gomez, Kaiser, and Polosukhin]{vaswani2017attention}
Vaswani, A.; Shazeer, N.; Parmar, N.; Uszkoreit, J.; Jones, L.; Gomez, A.N.;
  Kaiser, L.; Polosukhin, I.
\newblock Attention is All you Need.
\newblock In Proceedings of the Advances in Neural Information Processing
  Systems, December 4-9, 2017, Long Beach, CA, {USA},  2017, pp. 5998--6008.

\bibitem[Mohri \em{et~al.}(2002)Mohri, Pereira, and Riley]{mohri2002weighted}
Mohri, M.; Pereira, F.; Riley, M.
\newblock Weighted finite-state transducers in speech recognition.
\newblock {\em Computer Speech \& Language} {\bf 2002}, {\em 16},~69--88.

\bibitem[Mohri \em{et~al.}(2008)Mohri et~al.]{mohri2008speech}
Mohri, M.;  et~al.
\newblock Speech recognition with weighted finite-state transducers.
\newblock In Proceedings of the Springer Handbook of Speech Processing,  2008,
  pp. 559--584.

\bibitem[Riley \em{et~al.}(2009)Riley, Allauzen, and Jansche]{riley2009openfst}
Riley, M.; Allauzen, C.; Jansche, M.
\newblock {O}pen{F}st: An Open-Source, Weighted Finite-State Transducer Library
  and its Applications to Speech and Language.
\newblock In Proceedings of the Proceedings of Human Language Technologies: The
  2009 Annual Conference of the North {A}merican Chapter of the Association for
  Computational Linguistics, Companion Volume: Tutorial Abstracts; Association
  for Computational Linguistics: Boulder, Colorado,  2009; pp. 9--10.

\bibitem[Dahl \em{et~al.}(2011)Dahl, Yu, Deng, and Acero]{dahl2011context}
Dahl, G.E.; Yu, D.; Deng, L.; Acero, A.
\newblock Context-dependent pre-trained deep neural networks for
  large-vocabulary speech recognition.
\newblock {\em IEEE Transactions on audio, speech, and language processing}
  {\bf 2011}, {\em 20},~30--42.

\bibitem[Vesel{\'y} \em{et~al.}(2013)Vesel{\'y}, Ghoshal, Burget, and
  Povey]{vesely2013sequence}
Vesel{\'y}, K.; Ghoshal, A.; Burget, L.; Povey, D.
\newblock Sequence-discriminative training of deep neural networks.
\newblock In Proceedings of the Interspeech,  2013, Vol. 2013, pp. 2345--2349.

\bibitem[Morgan \em{et~al.}(1993)Morgan, Bourlard, Renals, Cohen, and
  Franco]{morgan1993hybrid}
Morgan, N.; Bourlard, H.; Renals, S.; Cohen, M.; Franco, H.
\newblock Hybrid neural network/hidden Markov model systems for continuous
  speech recognition. In {\em Advances in Pattern Recognition Systems Using
  Neural Network Technologies}; World Scientific,  1993; pp. 255--272.

\bibitem[Bourlard and Morgan(1993)]{bourlard1993connectionist}
Bourlard, H.A.; Morgan, N.
\newblock {\em Connectionist speech recognition: a hybrid approach}; Vol. 247,
  Springer Science \& Business Media,  1993.

\bibitem[Srinivasamurthy \em{et~al.}(2017)Srinivasamurthy, Motlicek, Himawan,
  Szaszak, Oualil, and Helmke]{srinivasamurthy2017semi}
Srinivasamurthy, A.; Motlicek, P.; Himawan, I.; Szaszak, G.; Oualil, Y.;
  Helmke, H.
\newblock Semi-supervised learning with semantic knowledge extraction for
  improved speech recognition in air traffic control.
\newblock In Proceedings of the Proc. of the 18th Annual Conference of the
  International Speech Communication Association,  2017.

\bibitem[Zuluaga-Gomez \em{et~al.}(2020)Zuluaga-Gomez, Motlicek, Zhan,
  Vesel{\`y}, and Braun]{zuluagagomez20_interspeech}
Zuluaga-Gomez, J.; Motlicek, P.; Zhan, Q.; Vesel{\`y}, K.; Braun, R.
\newblock {Automatic Speech Recognition Benchmark for Air-Traffic
  Communications}.
\newblock In Proceedings of the Interspeech,  2020, pp. 2297--2301.
\newblock {\url{https://doi.org/10.21437/Interspeech.2020-2173}}.

\bibitem[Zuluaga-Gomez \em{et~al.}(2021)Zuluaga-Gomez, Nigmatulina, Prasad,
  Motlicek, Vesel{\`y}, Kocour, and Sz{\"o}ke]{zuluagagomez21_interspeech}
Zuluaga-Gomez, J.; Nigmatulina, I.; Prasad, A.; Motlicek, P.; Vesel{\`y}, K.;
  Kocour, M.; Sz{\"o}ke, I.
\newblock {Contextual Semi-Supervised Learning: An Approach to Leverage
  Air-Surveillance and Untranscribed \MakeUppercase{ATC} Data in
  \MakeUppercase{ASR} Systems}.
\newblock In Proceedings of the Interspeech,  2021, pp. 3296--3300.
\newblock {\url{https://doi.org/10.21437/Interspeech.2021-1373}}.

\bibitem[Nigmatulina \em{et~al.}(2021)Nigmatulina, Braun, Zuluaga-Gomez, and
  Motlicek]{nigmatulina2021improving}
Nigmatulina, I.; Braun, R.; Zuluaga-Gomez, J.; Motlicek, P.
\newblock Improving callsign recognition with air-surveillance data in
  air-traffic communication.
\newblock {\em arXiv preprint arXiv:2108.12156} {\bf 2021}.

\bibitem[Graves and Jaitly(2014)]{graves2014towards}
Graves, A.; Jaitly, N.
\newblock Towards End-To-End Speech Recognition with Recurrent Neural Networks.
\newblock In Proceedings of the Proceedings of the 31th International
  Conference on Machine Learning, {ICML} 2014, Beijing, China, 21-26 June 2014.
  JMLR.org,  2014, Vol.~32, {\em {JMLR} Workshop and Conference Proceedings},
  pp. 1764--1772.

\bibitem[Chorowski \em{et~al.}(2015)Chorowski, Bahdanau, Serdyuk, Cho, and
  Bengio]{chorowski2015attention}
Chorowski, J.; Bahdanau, D.; Serdyuk, D.; Cho, K.; Bengio, Y.
\newblock Attention-Based Models for Speech Recognition.
\newblock In Proceedings of the Advances in Neural Information Processing
  Systems, December 7-12, 2015, Montreal, Quebec, Canada,  2015, pp. 577--585.

\bibitem[Schneider \em{et~al.}(2019)Schneider, Baevski, Collobert, and
  Auli]{schneider2019wav2vec}
Schneider, S.; Baevski, A.; Collobert, R.; Auli, M.
\newblock wav2vec: Unsupervised Pre-Training for Speech Recognition.
\newblock In Proceedings of the Interspeech 2019, Graz, Austria, 15-19
  September 2019. {ISCA},  2019, pp. 3465--3469.
\newblock {\url{https://doi.org/10.21437/Interspeech.2019-1873}}.

\bibitem[Chen \em{et~al.}(2021)Chen, Wang, Chen, et~al.]{chen2021wavlm}
Chen, S.; Wang, C.; Chen, Z.;  et~al.
\newblock Wavlm: Large-scale self-supervised pre-training for full stack speech
  processing.
\newblock {\em ArXiv preprint} {\bf 2021}, {\em abs/2110.13900}.

\bibitem[Oord \em{et~al.}(2018)Oord, Li, and Vinyals]{oord2018representation}
Oord, A.v.d.; Li, Y.; Vinyals, O.
\newblock Representation learning with contrastive predictive coding.
\newblock {\em ArXiv preprint} {\bf 2018}, {\em abs/1807.03748}.

\bibitem[Baevski \em{et~al.}(2020)Baevski, Schneider, and Auli]{baevski2019vq}
Baevski, A.; Schneider, S.; Auli, M.
\newblock vq-wav2vec: Self-Supervised Learning of Discrete Speech
  Representations.
\newblock In Proceedings of the 8th International Conference on Learning
  Representations, {ICLR} 2020, Addis Ababa, Ethiopia, April 26-30, 2020.
  OpenReview.net,  2020.

\bibitem[Zuluaga-Gomez \em{et~al.}(2023)Zuluaga-Gomez, Prasad, Nigmatulina,
  Sarfjoo, Motlicek, Kleinert, Helmke, Ohneiser, and Zhan]{zuluaga2022does}
Zuluaga-Gomez, J.; Prasad, A.; Nigmatulina, I.; Sarfjoo, S.; Motlicek, P.;
  Kleinert, M.; Helmke, H.; Ohneiser, O.; Zhan, Q.
\newblock How Does Pre-trained Wav2Vec2.0 Perform on Domain Shifted ASR? An
  Extensive Benchmark on Air Traffic Control Communications.
\newblock {\em IEEE Spoken Language Technology Workshop (SLT), Doha, Qatar}
  {\bf 2023}.

\bibitem[Yadav and Bethard(2018)]{yadav2018survey}
Yadav, V.; Bethard, S.
\newblock A Survey on Recent Advances in Named Entity Recognition from Deep
  Learning models.
\newblock In Proceedings of the Proceedings of the 27th International
  Conference on Computational Linguistics; Association for Computational
  Linguistics: Santa Fe, New Mexico, USA,  2018; pp. 2145--2158.

\bibitem[Sharma \em{et~al.}(2022)Sharma, Chakraborty, Kumar,
  et~al.]{sharma2022named}
Sharma, A.; Chakraborty, S.; Kumar, S.;  et~al.
\newblock Named Entity Recognition in Natural Language Processing: A Systematic
  Review.
\newblock In Proceedings of the Proceedings of Second Doctoral Symposium on
  Computational Intelligence. Springer,  2022, pp. 817--828.

\bibitem[Grishman and Sundheim(1996)]{grishman1996message}
Grishman, R.; Sundheim, B.
\newblock {M}essage {U}nderstanding {C}onference- 6: A Brief History.
\newblock In Proceedings of the {COLING} 1996 Volume 1: The 16th International
  Conference on Computational Linguistics,  1996.

\bibitem[Collobert \em{et~al.}(2011)Collobert, Weston, Bottou, Karlen,
  Kavukcuoglu, and Kuksa]{collobert2011natural}
Collobert, R.; Weston, J.; Bottou, L.; Karlen, M.; Kavukcuoglu, K.; Kuksa, P.
\newblock Natural language processing (almost) from scratch.
\newblock {\em Journal of machine learning research} {\bf 2011}, {\em
  12},~2493--2537.

\bibitem[Piskorski \em{et~al.}(2017)Piskorski, Pivovarova, {\v{S}}najder,
  Steinberger, and Yangarber]{piskorski2017first}
Piskorski, J.; Pivovarova, L.; {\v{S}}najder, J.; Steinberger, J.; Yangarber,
  R.
\newblock The First Cross-Lingual Challenge on Recognition, Normalization, and
  Matching of Named Entities in {S}lavic Languages.
\newblock In Proceedings of the Proceedings of the 6th Workshop on
  {B}alto-{S}lavic Natural Language Processing; Association for Computational
  Linguistics: Valencia, Spain,  2017; pp. 76--85.
\newblock {\url{https://doi.org/10.18653/v1/W17-1412}}.

\bibitem[Liu \em{et~al.}(2019)Liu, Ott, Goyal, Du, Joshi, Chen, Levy, Lewis,
  Zettlemoyer, and Stoyanov]{liu2019roberta}
Liu, Y.; Ott, M.; Goyal, N.; Du, J.; Joshi, M.; Chen, D.; Levy, O.; Lewis, M.;
  Zettlemoyer, L.; Stoyanov, V.
\newblock Roberta: A robustly optimized bert pretraining approach.
\newblock {\em ArXiv preprint} {\bf 2019}, {\em abs/1907.11692}.

\bibitem[He \em{et~al.}(2021)He, Liu, Gao, and Chen]{he2021deberta}
He, P.; Liu, X.; Gao, J.; Chen, W.
\newblock Deberta: decoding-Enhanced Bert with Disentangled Attention.
\newblock In Proceedings of the 9th International Conference on Learning
  Representations, {ICLR} 2021, Virtual Event, Austria, May 3-7, 2021.
  OpenReview.net,  2021.

\bibitem[Klatt(1987)]{klatt1987review}
Klatt, D.H.
\newblock Review of text-to-speech conversion for English.
\newblock {\em The Journal of the Acoustical Society of America} {\bf 1987},
  {\em 82},~737--793.

\bibitem[Murray \em{et~al.}(1996)Murray, Arnott, and
  Rohwer]{murray1996emotional}
Murray, I.R.; Arnott, J.L.; Rohwer, E.A.
\newblock Emotional stress in synthetic speech: Progress and future directions.
\newblock {\em Speech Communication} {\bf 1996}, {\em 20},~85--91.

\bibitem[Tokuda \em{et~al.}(2013)Tokuda, Nankaku, Toda, Zen, Yamagishi, and
  Oura]{tokuda2013speech}
Tokuda, K.; Nankaku, Y.; Toda, T.; Zen, H.; Yamagishi, J.; Oura, K.
\newblock Speech synthesis based on hidden Markov models.
\newblock {\em Proceedings of the IEEE} {\bf 2013}, {\em 101},~1234--1252.

\bibitem[Wang \em{et~al.}(2017)Wang, Skerry{-}Ryan, Stanton, Wu, Weiss, Jaitly,
  Yang, Xiao, Chen, Bengio, Le, Agiomyrgiannakis, Clark, and
  Saurous]{Wang2017tacotron}
Wang, Y.; Skerry{-}Ryan, R.J.; Stanton, D.; Wu, Y.; Weiss, R.J.; Jaitly, N.;
  Yang, Z.; Xiao, Y.; Chen, Z.; Bengio, S.;  et~al.
\newblock Tacotron: Towards End-to-End Speech Synthesis.
\newblock In Proceedings of the Interspeech 2017, Stockholm, Sweden, August
  20-24, 2017. {ISCA},  2017, pp. 4006--4010.

\bibitem[Shen \em{et~al.}(2018)Shen, Pang, Weiss, Schuster, Jaitly, Yang, Chen,
  Zhang, Wang, Ryan, Saurous, Agiomyrgiannakis, and Wu]{shen2018tacotron2}
Shen, J.; Pang, R.; Weiss, R.J.; Schuster, M.; Jaitly, N.; Yang, Z.; Chen, Z.;
  Zhang, Y.; Wang, Y.; Ryan, R.;  et~al.
\newblock Natural {TTS} Synthesis by Conditioning Wavenet on {MEL} Spectrogram
  Predictions.
\newblock In Proceedings of the 2018 {IEEE} International Conference on
  Acoustics, Speech and Signal Processing, {ICASSP} 2018, Calgary, AB, Canada,
  April 15-20, 2018. {IEEE},  2018, pp. 4779--4783.
\newblock {\url{https://doi.org/10.1109/ICASSP.2018.8461368}}.

\bibitem[He \em{et~al.}(2019)He, Deng, and He]{He2019}
He, M.; Deng, Y.; He, L.
\newblock {Robust Sequence-to-Sequence Acoustic Modeling with Stepwise
  Monotonic Attention for Neural TTS}.
\newblock In Proceedings of the Proc. Interspeech 2019,  2019, pp. 1293--1297.
\newblock {\url{https://doi.org/10.21437/Interspeech.2019-1972}}.

\bibitem[Kaur and Singh(2022)]{kaur2022conventional}
Kaur, N.; Singh, P.
\newblock Conventional and contemporary approaches used in text to speech
  synthesis: A review.
\newblock {\em Artificial Intelligence Review} {\bf 2022}, pp. 1--44.

\bibitem[Jeong \em{et~al.}(2021)Jeong, Kim, Cheon, Choi, and
  Kim]{jeong2021diff}
Jeong, M.; Kim, H.; Cheon, S.J.; Choi, B.J.; Kim, N.S.
\newblock Diff-tts: A denoising diffusion model for text-to-speech.
\newblock {\em arXiv preprint arXiv:2104.01409} {\bf 2021}.

\bibitem[Sarfjoo \em{et~al.}(2020)Sarfjoo, Madikeri, and
  Motlicek]{sarfjoo2020speech}
Sarfjoo, S.S.; Madikeri, S.; Motlicek, P.
\newblock Speech activity detection based on multilingual speech recognition
  system.
\newblock {\em arXiv preprint arXiv:2010.12277} {\bf 2020}.

\bibitem[Ariav and Cohen(2019)]{ariav2019end}
Ariav, I.; Cohen, I.
\newblock An end-to-end multimodal voice activity detection using wavenet
  encoder and residual networks.
\newblock {\em IEEE Journal of Selected Topics in Signal Processing} {\bf
  2019}, {\em 13},~265--274.

\bibitem[Ding \em{et~al.}(2019)Ding, Wang, Chang, Wan, and
  Moreno]{ding2019personal}
Ding, S.; Wang, Q.; Chang, S.y.; Wan, L.; Moreno, I.L.
\newblock Personal VAD: Speaker-conditioned voice activity detection.
\newblock {\em arXiv preprint arXiv:1908.04284} {\bf 2019}.

\bibitem[Medennikov \em{et~al.}(2020)Medennikov, Korenevsky, Prisyach,
  Khokhlov, Korenevskaya, Sorokin, Timofeeva, Mitrofanov, Andrusenko,
  Podluzhny, et~al.]{medennikov2020target}
Medennikov, I.; Korenevsky, M.; Prisyach, T.; Khokhlov, Y.; Korenevskaya, M.;
  Sorokin, I.; Timofeeva, T.; Mitrofanov, A.; Andrusenko, A.; Podluzhny, I.;
  et~al.
\newblock Target-speaker voice activity detection: a novel approach for
  multi-speaker diarization in a dinner party scenario.
\newblock {\em arXiv preprint arXiv:2005.07272} {\bf 2020}.

\bibitem[Zazo \em{et~al.}(2016)Zazo, Sainath, Simko, and Parada]{Zazo_2016}
Zazo, R.; Sainath, T.N.; Simko, G.; Parada, C.
\newblock Feature Learning with Raw-Waveform CLDNNs for Voice Activity
  Detection.
\newblock In Proceedings of the Interspeech 2016,  2016, pp. 3668--3672.
\newblock {\url{https://doi.org/10.21437/Interspeech.2016-268}}.

\bibitem[Ng \em{et~al.}(2012)Ng, Zhang, Nguyen, Matsoukas, Zhou, Mesgarani,
  Vesel{\`y}, and Mat{\v{e}}jka]{ng2012developing}
Ng, T.; Zhang, B.; Nguyen, L.; Matsoukas, S.; Zhou, X.; Mesgarani, N.;
  Vesel{\`y}, K.; Mat{\v{e}}jka, P.
\newblock Developing a speech activity detection system for the DARPA RATS
  program.
\newblock In Proceedings of the Thirteenth annual conference of the
  international speech communication association,  2012.

\bibitem[Chang \em{et~al.}(2018)Chang, Li, Simko, Sainath, Tripathi, van~den
  Oord, and Vinyals]{chang2018temporal}
Chang, S.Y.; Li, B.; Simko, G.; Sainath, T.N.; Tripathi, A.; van~den Oord, A.;
  Vinyals, O.
\newblock Temporal modeling using dilated convolution and gating for
  voice-activity-detection.
\newblock In Proceedings of the 2018 IEEE international conference on
  acoustics, speech and signal processing (ICASSP). IEEE,  2018, pp.
  5549--5553.

\bibitem[Helmke \em{et~al.}(2022)Helmke, Ondřej, Shetty, Arilíusson,
  Simiganoschi, Kleinert, Ohneiser, Ehr, Zuluaga-Gomez, and
  Smrz]{hartmut2022readback}
Helmke, H.; Ondřej, K.; Shetty, S.; Arilíusson, H.; Simiganoschi, T.;
  Kleinert, M.; Ohneiser, O.; Ehr, H.; Zuluaga-Gomez, J.; Smrz, P.
\newblock {Readback Error Detection by Automatic Speech Recognition and
  Understanding -- Results of HAAWAII Project for Isavia’s Enroute Airspace}.
\newblock In Proceedings of the 12th SESAR Innovation Days. Sesar Joint
  Undertaking.,  2022.

\bibitem[Cordero \em{et~al.}(2012)Cordero, Dorado, and
  de~Pablo]{cordero2012automated}
Cordero, J.M.; Dorado, M.; de~Pablo, J.M.
\newblock Automated speech recognition in {ATC} environment.
\newblock In Proceedings of the Proceedings of the 2nd International Conference
  on Application and Theory of Automation in Command and Control Systems,
  2012, pp. 46--53.

\bibitem[Delpech \em{et~al.}(2018)Delpech, Laignelet, Pimm, Raynal, Trzos,
  Arnold, and Pronto]{AIRBUS}
Delpech, E.; Laignelet, M.; Pimm, C.; Raynal, C.; Trzos, M.; Arnold, A.;
  Pronto, D.
\newblock {A Real-life, French-accented Corpus of Air Traffic Control
  Communications}.
\newblock In Proceedings of the Proceedings of the Eleventh International
  Conference on Language Resources and Evaluation (LREC 2018),  2018.

\bibitem[Segura \em{et~al.}(2007)Segura, Ehrette, Potamianos, Fohr, Illina,
  Breton, Clot, Gemello, Matassoni, and Maragos]{HIWIRE}
Segura, J.; Ehrette, T.; Potamianos, A.; Fohr, D.; Illina, I.; Breton, P.;
  Clot, V.; Gemello, R.; Matassoni, M.; Maragos, P.
\newblock The {HIWIRE} database, a noisy and non-native {English} speech corpus
  for cockpit communication.
\newblock {\em Online. http://www. hiwire. org} {\bf 2007}.

\bibitem[Gulati \em{et~al.}(2020)Gulati, Qin, Chiu, Parmar, Zhang, Yu, Han,
  Wang, Zhang, Wu, and Pang]{Gulati2020-conformer}
Gulati, A.; Qin, J.; Chiu, C.C.; Parmar, N.; Zhang, Y.; Yu, J.; Han, W.; Wang,
  S.; Zhang, Z.; Wu, Y.;  et~al.
\newblock {Conformer: Convolution-augmented Transformer for Speech
  Recognition}.
\newblock In Proceedings of the Proc. Interspeech 2020,  2020, pp. 5036--5040.
\newblock {\url{https://doi.org/10.21437/Interspeech.2020-3015}}.

\bibitem[LeCun \em{et~al.}(1995)LeCun, Bengio, et~al.]{lecun1995convolutional}
LeCun, Y.; Bengio, Y.;  et~al.
\newblock Convolutional networks for images, speech, and time series.
\newblock {\em The handbook of brain theory and neural networks} {\bf 1995},
  {\em 3361},~1995.

\bibitem[Povey \em{et~al.}(2018)Povey, Cheng, Wang, Li, Xu, Yarmohammadi, and
  Khudanpur]{povey2018semi}
Povey, D.; Cheng, G.; Wang, Y.; Li, K.; Xu, H.; Yarmohammadi, M.; Khudanpur, S.
\newblock Semi-Orthogonal Low-Rank Matrix Factorization for Deep Neural
  Networks.
\newblock In Proceedings of the Interspeech,  2018, pp. 3743--3747.

\bibitem[Povey \em{et~al.}(2011)Povey, Ghoshal, Boulianne, Burget, Glembek,
  Goel, Hannemann, Motlicek, Qian, Schwarz, et~al.]{povey2011kaldi}
Povey, D.; Ghoshal, A.; Boulianne, G.; Burget, L.; Glembek, O.; Goel, N.;
  Hannemann, M.; Motlicek, P.; Qian, Y.; Schwarz, P.;  et~al.
\newblock {The Kaldi speech recognition toolkit}.
\newblock In Proceedings of the IEEE workshop on automatic speech recognition
  and understanding. IEEE Signal Processing Society,  2011, number CONF.

\bibitem[Vyas \em{et~al.}(2021)Vyas, Madikeri, and Bourlard]{9414741}
Vyas, A.; Madikeri, S.; Bourlard, H.
\newblock Lattice-Free Mmi Adaptation of Self-Supervised Pretrained Acoustic
  Models.
\newblock In Proceedings of the ICASSP 2021 - 2021 IEEE International
  Conference on Acoustics, Speech and Signal Processing (ICASSP),  2021, pp.
  6219--6223.
\newblock {\url{https://doi.org/10.1109/ICASSP39728.2021.9414741}}.

\bibitem[Babu \em{et~al.}(2021)Babu, Wang, Tjandra, Lakhotia, Xu, Goyal, Singh,
  von Platen, Saraf, Pino, et~al.]{babu2021xls}
Babu, A.; Wang, C.; Tjandra, A.; Lakhotia, K.; Xu, Q.; Goyal, N.; Singh, K.;
  von Platen, P.; Saraf, Y.; Pino, J.;  et~al.
\newblock Xls-r: Self-supervised cross-lingual speech representation learning
  at scale.
\newblock {\em ArXiv preprint} {\bf 2021}, {\em abs/2111.09296}.

\bibitem[Ravanelli \em{et~al.}(2021)Ravanelli, Parcollet, Plantinga, Rouhe,
  Cornell, Lugosch, Subakan, Dawalatabad, Heba, Zhong,
  et~al.]{ravanelli2021speechbrain}
Ravanelli, M.; Parcollet, T.; Plantinga, P.; Rouhe, A.; Cornell, S.; Lugosch,
  L.; Subakan, C.; Dawalatabad, N.; Heba, A.; Zhong, J.;  et~al.
\newblock SpeechBrain: A general-purpose speech toolkit.
\newblock {\em arXiv preprint arXiv:2106.04624} {\bf 2021}.

\bibitem[Panayotov \em{et~al.}(2015)Panayotov, Chen, Povey, and
  Khudanpur]{panayotov_librispeech_ICASSP2015}
Panayotov, V.; Chen, G.; Povey, D.; Khudanpur, S.
\newblock Librispeech: an {ASR} corpus based on public domain audio books.
\newblock In Proceedings of the International Conference on Acoustics, Speech
  and Signal Processing (ICASSP). IEEE,  2015, pp. 5206--5210.

\bibitem[Srivastava \em{et~al.}(2014)Srivastava, Hinton, Krizhevsky, Sutskever,
  and Salakhutdinov]{srivastava2014dropout}
Srivastava, N.; Hinton, G.; Krizhevsky, A.; Sutskever, I.; Salakhutdinov, R.
\newblock Dropout: a simple way to prevent neural networks from overfitting.
\newblock {\em The journal of machine learning research} {\bf 2014}, {\em
  15},~1929--1958.

\bibitem[Hendrycks and Gimpel(2016)]{hendrycks2016gaussian}
Hendrycks, D.; Gimpel, K.
\newblock Gaussian error linear units (gelus).
\newblock {\em ArXiv preprint} {\bf 2016}, {\em abs/1606.08415}.

\bibitem[Kingma and Ba(2014)]{kingma2014adam}
Kingma, D.P.; Ba, J.
\newblock Adam: A method for stochastic optimization.
\newblock {\em arXiv preprint arXiv:1412.6980} {\bf 2014}.

\bibitem[Karita \em{et~al.}(2019)Karita, Chen, Hayashi, Hori, Inaguma, Jiang,
  Someki, Soplin, Yamamoto, Wang, et~al.]{karita2019comparative}
Karita, S.; Chen, N.; Hayashi, T.; Hori, T.; Inaguma, H.; Jiang, Z.; Someki,
  M.; Soplin, N.E.Y.; Yamamoto, R.; Wang, X.;  et~al.
\newblock A comparative study on transformer vs rnn in speech applications.
\newblock In Proceedings of the 2019 IEEE Automatic Speech Recognition and
  Understanding Workshop (ASRU). IEEE,  2019, pp. 449--456.

\bibitem[Graves and Graves(2012)]{ctc_loss}
Graves, A.; Graves, A.
\newblock Connectionist temporal classification.
\newblock {\em Supervised sequence labelling with recurrent neural networks}
  {\bf 2012}, pp. 61--93.

\bibitem[Chen \em{et~al.}(2019)Chen, Jain, Wang, Seltzer, and
  Fuegen]{chen2019end}
Chen, Z.; Jain, M.; Wang, Y.; Seltzer, M.L.; Fuegen, C.
\newblock End-to-end contextual speech recognition using class language models
  and a token passing decoder.
\newblock In Proceedings of the ICASSP 2019-2019 IEEE International Conference
  on Acoustics, Speech and Signal Processing (ICASSP). IEEE,  2019, pp.
  6186--6190.

\bibitem[Wolf \em{et~al.}(2020)Wolf, Debut, Sanh, Chaumond, Delangue, Moi,
  Cistac, Rault, Louf, Funtowicz, et~al.]{wolf2020transformers}
Wolf, T.; Debut, L.; Sanh, V.; Chaumond, J.; Delangue, C.; Moi, A.; Cistac, P.;
  Rault, T.; Louf, R.; Funtowicz, M.;  et~al.
\newblock Transformers: State-of-the-Art Natural Language Processing.
\newblock In Proceedings of the Proceedings of the 2020 Conference on Empirical
  Methods in Natural Language Processing: System Demonstrations,  2020, pp.
  38--45.

\bibitem[Lhoest \em{et~al.}(2021)Lhoest, del Moral, Jernite, Thakur, von
  Platen, Patil, Chaumond, Drame, Plu, Tunstall, et~al.]{lhoest2021datasets}
Lhoest, Q.; del Moral, A.V.; Jernite, Y.; Thakur, A.; von Platen, P.; Patil,
  S.; Chaumond, J.; Drame, M.; Plu, J.; Tunstall, L.;  et~al.
\newblock Datasets: A Community Library for Natural Language Processing.
\newblock In Proceedings of the Proceedings of the 2021 Conference on Empirical
  Methods in Natural Language Processing: System Demonstrations,  2021, pp.
  175--184.

\bibitem[Pascanu \em{et~al.}(2013)Pascanu, Mikolov, and
  Bengio]{pascanu2013difficulty}
Pascanu, R.; Mikolov, T.; Bengio, Y.
\newblock On the difficulty of training recurrent neural networks.
\newblock In Proceedings of the Proceedings of the 30th International
  Conference on Machine Learning, {ICML} 2013, Atlanta, GA, USA, 16-21 June
  2013. JMLR.org,  2013, Vol.~28, {\em {JMLR} Workshop and Conference
  Proceedings}, pp. 1310--1318.

\bibitem[Loshchilov and Hutter(2019)]{loshchilov2018decoupled}
Loshchilov, I.; Hutter, F.
\newblock Decoupled Weight Decay Regularization.
\newblock In Proceedings of the 7th International Conference on Learning
  Representations, {ICLR} 2019, New Orleans, LA, USA, May 6-9, 2019.
  OpenReview.net,  2019.

\bibitem[Rigault \em{et~al.}(2022)Rigault, Cevenini, Choukri, Kocour,
  Vesel{\`y}, Szoke, Motlicek, Zuluaga-Gomez, Blatt, Klakow,
  et~al.]{rigault2022legal}
Rigault, M.; Cevenini, C.; Choukri, K.; Kocour, M.; Vesel{\`y}, K.; Szoke, I.;
  Motlicek, P.; Zuluaga-Gomez, J.P.; Blatt, A.; Klakow, D.;  et~al.
\newblock Legal and ethical challenges in recording air traffic control speech.
\newblock In Proceedings of the Proceedings of the Workshop on Ethical and
  Legal Issues in Human Language Technologies and Multilingual
  De-Identification of Sensitive Data In Language Resources within the 13th
  Language Resources and Evaluation Conference,  2022, pp. 79--83.

\end{thebibliography}

\end{document}